\documentclass{elsart}

\usepackage{graphicx}
\usepackage{amsmath}
\usepackage{bm}  
\usepackage[titles]{tocloft}

\journal{Annals of Physics}

\newcommand{\beq}{\begin{eqnarray}}
\newcommand{\eeq}{\end{eqnarray}}
\newcommand{\bqa}{\begin{eqnarray}}
\newcommand{\eqa}{\end{eqnarray}}

\newcommand{\gsim}{\hspace*{0.2em}\raisebox{0.5ex}{$>$}
     \hspace{-0.8em}\raisebox{-0.3em}{$\sim$}\hspace*{0.2em}}
\newcommand{\lsim}{\hspace*{0.2em}\raisebox{0.5ex}{$<$}
     \hspace{-0.8em}\raisebox{-0.3em}{$\sim$}\hspace*{0.2em}}
\def\mqo2{{\!\!\!}}

\begin{document}
\begin{frontmatter}

\title{Efimov Physics in Cold Atoms}

\author[OSU]{Eric Braaten}
\ead{braaten@mps.ohio-state.edu}
\address[OSU]{Department of Physics, The Ohio State University,
                Columbus, OH 43210, USA}
\author[Bonn]{H.-W. Hammer}
\ead{hammer@itkp.uni-bonn.de}
\address[Bonn]{Helmholtz-Institut f\"ur Strahlen- und Kernphysik (Theorie),
 Universit\"at Bonn, 53115 Bonn, Germany}

\date{\today}

\begin{abstract}
Atoms with a large scattering length 
have universal low-energy properties that do not depend on the details 
of their structure or their interactions at short distances.
In the 2-atom sector, the universal properties are familiar and 
depend only on the scattering length.  
In the 3-atom sector for identical bosons, the universal properties 
include the existence of a sequence of shallow triatomic molecules 
called {\it Efimov trimers} 
and log-periodic dependence of scattering observables
on the energy and the scattering length. 
In this review, we summarize the universal results that are 
currently known.  
We also summarize the experimental information 
that is currently available with an emphasis on 3-atom loss processes.
\end{abstract}

%\maketitle
%\tableofcontents
%\thispagestyle{empty}

\end{frontmatter}

%\newpage
%\tableofcontents

%%%%%%%%%%%%%%%%%%%%%%%%%%%%%%%%%%%%%%%%%%%%%%%%%%
%
%     INTRODUCTION
%
%%%%%%%%%%%%%%%%%%%%%%%%%%%%%%%%%%%%%%%%%%%%%%%%%%

\newpage
\section{Introduction}
        \label{sec:intro}

The scattering of two particles with
sufficiently low kinetic energy is determined by their 
S-wave {\it scattering length}, which is commonly denoted by $a$.
The energies of the particles are sufficiently low if their de Broglie
wavelengths are large compared to the range of the
interaction. The scattering length $a$ is also the most important 
interaction variable for 3-body systems and for higher 
$N$-body systems if all their constituents have sufficiently 
low energy.  

Generically, the scattering length $a$ is comparable in magnitude 
to the range $r_0$ of the interaction:  $|a| \sim r_0$.
In exceptional cases, the scattering length can be much larger 
in magnitude than the range:  $|a| \gg r_0$.
Such a {\it large scattering length} requires the fine-tuning
of a parameter characterizing the interactions to 
the neighborhood of a critical value at which $a$ diverges
to $\pm \infty$.  If the scattering length is large, the atoms 
exhibit properties that depend on $a$ but are insensitive 
to the range and other details of the short-range interaction.
We will refer to these properties as {\it universal},
because they apply equally well to any nonrelativistic particle 
with short range interactions that produce a large scattering length.

In the 2-body sector, the universal properties 
are simple but nontrivial.  For example, in the case of 
equal-mass particles with $a>0$, there is a 2-body bound state 
near the scattering threshold with binding energy
$E_D  =  \hbar^2 /(m a^2)$.
The corrections to this formula are parametrically small: 
they are suppressed by powers of $r_0/a$. 

In the 3-body sector, the universal properties 
are much more intricate.
The first strong evidence for universal behavior in the 3-body system 
was the discovery of the {\it Efimov effect} 
by Vitaly Efimov in 1969 \cite{Efimov70}.
In the {\it resonant limit} $a \to \pm \infty$, there is a 2-body bound state 
exactly at the 2-body scattering threshold $E = 0$.  
Efimov showed that in this limit there can also be infinitely many,
arbitrarily-shallow 3-body bound states whose binding energies 
$E_T^{(n)}$ have an accumulation point at $E = 0$.
They are called {\it Efimov states}.  
As the threshold is approached, the ratio of the binding energies 
of successive Efimov states approaches a universal constant.
In the case of identical bosons, the asymptotic ratio is
%----------------------
\begin{eqnarray}
E_T^{(n+1)}/E_T^{(n)} & \longrightarrow & 1/515.03\,, 
%\nonumber \\ && 
\qquad
{\rm as \ } n \to + \infty {\rm \ \ with \ } a = \pm \infty \,.
\label{R-resonant}
\end{eqnarray}
%----------------------
%[E,H]
The universal ratio in Eq.~(\ref{R-resonant}) is independent of the mass 
or structure of the identical
particles and independent of the form of their short-range interactions. 
The Efimov effect can also occur in other 3-body systems if at least two 
of the three pairs have a large S-wave scattering length.
The numerical value of the asymptotic ratio
may differ from the value in Eq.~(\ref{R-resonant}).  
Efimov's discovery was at first greeted with some skepticism.
However the theoretical evidence for the Efimov effect 
quickly became conclusive.

The Efimov effect proved to be just the first nugget from a gold mine
of universal aspects of the 3-body problem.
This system has universal properties not only in
the resonant limit $a = \pm \infty$,
but whenever the scattering length is large
compared to the range $r_0$.
In two brilliant papers in 1971 and 1979 \cite{Efimov71,Efimov79}, 
Efimov derived a number of universal results on  
low-energy 3-body observables for three identical bosons.
The dependence of these results on the scattering length 
or the energy is
characterized by scaling behavior modulo coefficients 
that are log-periodic functions.
This behavior is characteristic of a system with a
{\it discrete scaling symmetry}. 
We will refer to universal aspects associated with a 
discrete scaling symmetry as {\it Efimov physics}.

Although the existence of Efimov states quickly became
well-established theoretically, their experimental confirmation
has proved to be more challenging.
Although more than 36 years have elapsed since Efimov's discovery,
there has still not been any convincing direct observation 
of an Efimov state.
One promising system for observing Efimov states is
$^4$He atoms, which have a scattering length that is more than 
a factor of 10 larger than the range of the interaction.
Calculations using accurate potential models indicate that the
system of three $^4$He atoms has two 3-body bound states or trimers.
The ground-state trimer is interpreted by some (including the authors)
as an Efimov state, and it has been observed in experiments 
involving the scattering of cold jets of $^4$He atoms 
from a diffraction grating \cite{STo96}. 
The excited trimer is universally believed to be an Efimov state, 
but it has not yet been observed.

The rapid development of the field of cold atom physics has opened
up new opportunities for the experimental study of Efimov physics.
This is made possible by two separate technological developments.
One is the technology for cooling atoms to the extremely low temperatures
where Efimov physics plays a dramatic role.
The other is the technology for controlling the interactions 
between atoms.  
By tuning the magnetic field to a Feshbach resonance,
the scattering lengths of the atoms can be made arbitrarily large.
Both of these technological developments were crucial in a recent 
experiment that provided the first indirect evidence for the 
existence of an Efimov state \cite{Kraemer06}.
The signature of the Efimov state was a resonant enhancement
of the loss rate from 3-body recombination
in an ultracold gas of $^{133}$Cs atoms. 

This experiment is just the beginning of the study 
of Efimov physics in ultracold atoms.  We have recently written 
a thorough review of universality in few-body physics 
with large scattering length \cite{bigrev}.
In this shorter review, we summarize the results derived in 
Ref.~\cite{bigrev}, include a few more recent developments, 
and focus more directly on applications in atomic physics.

In Section~\ref{sec:scat}, we introduce various scattering concepts 
that play an important role in Efimov physics.
Most of this review is focused on identical bosons, 
because this is the simplest case in which Efimov states arise
and because it is the system for which  
Efimov physics has been most thoroughly explored.  
In Section~\ref{sec:uni2} and \ref{sec:uni3}, 
we summarize the universal features of 
identical bosons with a large scattering length
in the 2-body and 3-body sectors, respectively.
In Section~\ref{sec:deep}, we describe the effects of deep diatomic
molecules on the universal results. 
In Section~\ref{sec:appatom}, we describe applications to 
$^4$He atoms and to alkali atoms near a Feshbach resonance.
In Section~\ref{sec:beyond}, we discuss the conditions under which
Efimov physics arises in systems other than identical bosons.
In Section~\ref{sec:beyondsl}, we consider the corrections to 
the universal results from the nonzero effective range
and from microscopic models of the atoms as well as the
extension of universality to 4-body systems.
We conclude with the outlook for the study of Efimov physics 
in ultracold atoms.

%\newpage

%%%%%%%%%%%%%%%%%%%%%%%%%%%%%%%%%%%%%%%%%%%%%%%%%%
%
%     SCATTERING CONCEPTS
%
%%%%%%%%%%%%%%%%%%%%%%%%%%%%%%%%%%%%%%%%%%%%%%%%%%

\section{Scattering Concepts}
        \label{sec:scat}

%%%%%%%%%%%%%%%%%%%%%%%%%%%%%%%%%%%%%%%%%%%%%%%%%%
%   Scattering length
%%%%%%%%%%%%%%%%%%%%%%%%%%%%%%%%%%%%%%%%%%%%%%%%%%

\subsection{Scattering length}

The elastic scattering of two atoms of mass $m$ 
and total kinetic energy $E = \hbar^2 k^2/m$ 
can be described by a stationary wave function $\psi({\bm r})$
that depends on the separation vector ${\bm r}$ of the two atoms.
Its asymptotic behavior as $r \to \infty$ is the sum of a 
plane wave and an outgoing spherical wave:
%----------------------
\begin{eqnarray}
\psi({\bm r}) \longrightarrow e^{i k z} + f_k(\theta) {e^{ikr} \over r} \,,
\label{psi-f}
\end{eqnarray}
%----------------------
%[E,H]
where $f_k (\theta)$ is the {\it scattering amplitude},
which depends on the scattering angle $\theta$ and the wave number $k$.
The {\it differential cross section} $d \sigma/d \Omega$ 
for the scattering of identical bosons can be 
expressed in the form
%----------------------
\begin{eqnarray}
\frac{d \sigma}{d \Omega} = 
\left|f_k (\theta) + f_k (\pi - \theta)\right|^2 \,.
\label{dcross}
\end{eqnarray}
%----------------------
%[H,E] 
If the two atoms are identical fermions, the $+$ should be replaced by $-$.
If they are distinguishable atoms, 
the term $+ f_k (\pi-\theta)$ should be omitted.
The elastic cross section $\sigma(E)$ is obtained by integrating 
over only ${1 \over 2}$ the $4 \pi$ solid angle 
if the two atoms are identical bosons or identical fermions
and over the entire $4 \pi$ solid angle if they are distinguishable.

The {\it partial-wave expansion} resolves the scattering amplitude $f_k 
(\theta)$ into contributions from definite angular momentum quantum number 
$L$ by expanding it in terms of Legendre polynomials of $\cos \theta$:
%----------------------
\begin{eqnarray}
f_k (\theta) = \sum_{L=0}^\infty
{ 2L+1 \over k \cot \delta_L(k) - ik} P_L (\cos \theta) \,,
\label{pwe}
\end{eqnarray}
%----------------------
%[H,E] 
If there are no inelastic 2-body channels, 
the {\it phase shifts} $\delta_L(k)$ are real-valued.
If there are inelastic channels, the phase shifts can be
complex-valued with positive imaginary parts.

If the atoms interact through a short-range 2-body potential, then
the phase shift $\delta_L (k)$ approaches zero like $k^{2L+1}$
in the low-energy limit $k \to 0$.
Thus S-wave ($L=0)$ scattering dominates in the low-energy limit
unless the atoms are identical fermions, 
in which case P-wave ($L=1)$ scattering dominates.
At sufficiently low energies, the S-wave phase shift
$\delta_0 (k)$ can be expanded in powers of $k^2$
\cite{Schwinger47}. The expansion is
called the {\it effective-range expansion} and is
conventionally expressed in the form
%----------------------
\begin{eqnarray}
k \cot \delta_0 (k) = - 1/a + \mbox{$1\over 2$} r_s k^2 + \ldots \,,
\label{kcot}
\end{eqnarray}
%----------------------
%[H,E] 
where $a$ is the scattering length and $r_s$ the
S-wave {\it effective range}.

%%%%%%%%%%%%%%%%%%%%%%%%%%%%%%%%%%%%%%%%%%%%%%%%%%  
%    Natural low-energy length scale
%%%%%%%%%%%%%%%%%%%%%%%%%%%%%%%%%%%%%%%%%%%%%%%%%%

\subsection{Natural low-energy length scale}
        \label{subsec:natural}

At sufficiently low energies, atoms behave like point particles with 
short-range interactions.
The length scale that governs the quantum behavior of the center-of-mass 
coordinate of an atom is the de Broglie wavelength $\lambda = 2 \pi \hbar/p$,
where $p$ is the momentum of the atom.
If the relative momentum $p$ of two atoms is sufficiently small,
their de Broglie wavelengths are larger than the spacial extent of the
atoms and they are unable to resolve each other's internal structure.  
Their interactions will therefore be indistinguishable 
from those of point particles.
If the atoms interact through a short-range potential
with range $r_0$ and if the relative momentum of the two atoms
satisfies $p \ll  \hbar/ r_0$,
then their de Broglie wavelengths prevent
them from resolving the structure of the potential.

For real atoms, the potential is not quite short-range.
The interatomic potential $V(r)$ between two neutral atoms 
in their ground states consists of a short range potential and
a long-range tail provided by the van der Waals interaction:
%----------------------
\begin{eqnarray}
V(r) \longrightarrow - {C_6 \over r^6}
\qquad \mbox{ as\ }  r \to \infty \,.
\label{V-vdW}
\end{eqnarray}
%----------------------
%[H,E] 
The van der Waals tail of the potential does not prevent 
the scattering amplitude from being expanded in powers of the 
relative momentum ${\bm k}$, but at 4$^{\rm th}$ order the dependence 
on ${\bm k}$ becomes nonpolynomial.
The coefficient $C_6$ determines a length scale 
called the van der Waals length:
%----------------------
\begin{eqnarray}
\ell_{\rm vdW} = \left(m C_6/\hbar^2 \right)^{1/4} \,.
\label{ellvdW}
\end{eqnarray}
%----------------------
%[H,E] 
At sufficiently low energy, the interactions between atoms 
are dominated by the van der Waals interaction.
Thus the van der Waals length in Eq.~(\ref{ellvdW})
is the natural low-energy length scale for atoms.

The natural low-energy length scale sets the natural scale for the
coefficients in the low-energy expansion 
of the scattering amplitude $f_k(\theta)$.
It is sometimes referred to as the 
{\it characteristic radius of interaction} and often denoted by $r_0$. 
If the magnitude $|a|$ of the scattering length is comparable to $\ell$, 
we say that $a$ has a {\it natural size}.
If $|a| \gg \ell$, we call the scattering length {\it unnaturally large},
or just {\it large} to be concise.
The natural low-energy length scale also sets the natural scale for 
the effective range $r_s$ defined by Eq.~(\ref{kcot}). 
Even if $a$ is large, we expect $r_s$ to have a natural
magnitude of order $\ell$.  For $a$ and $r_s$ to both be unnaturally large
would require the simultaneous fine-tuning of two parameters in the potential.

The natural low-energy length scale $\ell$ also sets the natural scale for the
binding energies of the 2-body bound states closest to threshold. The
binding energy of the shallowest bound state is expected to be
of order $\hbar^2 /m \ell^2$ or larger. 
It can be orders of magnitude smaller only if there is
a large positive scattering length $a \gg \ell$.

%%%%%%%%%%%%%%%%%%%%%%%%%%%%%%%%%%%%%%%%%%%%%%%%%%
%   Atoms with large scattering length
%%%%%%%%%%%%%%%%%%%%%%%%%%%%%%%%%%%%%%%%%%%%%%%%%%

\subsection{Atoms with large scattering length}
\label{subsec:atomexamp}

The scattering length $a$ can be orders of magnitude larger 
than the natural low-energy scale only if some parameter is tuned 
to near a critical value at which $a$ diverges.
This fine-tuning can be 
due to fortuitous values of the fundamental constants of nature,
in which case we call it {\it accidental fine-tuning},
or it can be due to the adjustment of parameters 
that are under experimental control,
in which case we call it {\it experimental fine-tuning}.  
We will give examples of atoms with both kinds of fine-tunings.

The simplest example of an atom with a large positive scattering length 
is the helium atom $^4$He.  
The van der Waals length defined by Eq.~(\ref{ellvdW}) is 
$\ell_{\rm v dW} \approx 10.2 \, a_0$, where 
$a_0=0.529177$ \AA\  is the Bohr radius.
The scattering length can be calculated precisely using 
potential models for helium atoms.  
For example, the LM2M2 \cite{AS91} and TTY \cite{TTY95} 
potentials have a large scattering length 
$a = 189 \, a_0$ but a natural effective range $r_s = 14 \, a_0$. 
They predict that $^4$He atoms have a single 2-body bound state 
or {\it dimer}, which is very weakly bound. 
The binding energy of the dimer is $E_2= 1.31$ mK,
which is much smaller than the natural low-energy scale 
$\hbar^2/m \ell_{\rm vdW}^2 \approx 420$ mK\footnote{
The conversion formula for this energy unit is
1 mK = $8.61734 \times 10^{-8}$ eV.}.
The scattering length of $^4$He atoms is large because of an 
{\it accidental fine-tuning}.  The mass of the $^4$He nucleus, 
the electron mass, and the fine structure constant $\alpha$ of QED 
have fortuitous values that make the potential between two $^4$He atoms 
just deep enough to have a bound state very close to threshold, 
and therefore a large scattering length.  If one of the $^4$He atoms 
is replaced by a $^3$He atom, which decreases the reduced mass by 14\% 
without changing the interaction potential, 
the scattering length has the more natural value $-33 \, a_0$.

The simplest example of an atom with a large negative scattering length 
is the polarized tritium atom $^3$H \cite{BEGKH02}.  
The van der Waals length for $^3$H is $\ell_{\rm vdW} = 13.7 \, a_0$. 
The scattering length for polarized $^3$H atoms is the
spin-triplet scattering length $a_t = -82.1 \, a_0$ \cite{BEGKH02},
which is much larger than $\ell_{\rm vdW}$. 
Polarized tritium atoms have no 2-body bound states,
but they have a single 3-body bound state 
with a shallow binding energy of about 4.59 mK \cite{BEGKH02}. 

Other examples of atoms with large scattering lengths due to 
accidental fine tuning can be found among the heavier alkali atoms.  
The spin-triplet scattering lengths $a_t$ for $^6$Li and for $^{133}$Cs
and the spin-singlet scattering length $a_s$ for $^{85}$Rb
are all more than an order of magnitude larger than the 
corresponding van der Waals scales $\ell_{\rm vdW}$. 
The fine-tuning is illustrated by the facts that $^7$Li,
whose mass is 17\% larger than  that of $^6$Li, has a natural value for $a_t$
and that $^{87}$Rb, whose mass is 2.3\% larger than  
that of $^{85}$Rb, has a natural value for $a_s$. 

The mechanism for generating a large scattering length that involves 
tuning the depth or range of the potential is called a 
{\it shape resonance}.
With this mechanism, only the {\it open channel} defined by the 
scattering particles plays an important role.
Another mechanism for generating a  large scattering length is a
{\it Feshbach resonance}  \cite{Feshbach62}.
This requires a second {\it closed channel} in which
scattering states are energetically forbidden
that is weakly coupled to the open channel.  
A large scattering length for particles in the open channel
can be generated by tuning the depth of the potential for the closed 
channel to bring one of its bound states close to the threshold for the 
open channel. 
The resulting enhancement of the scattering of particles 
in the open channel is a Feshbach resonance.

Feshbach resonances in alkali atoms can be created by tuning the 
magnetic field \cite{TMVS92,TVS93}.
In this case, the open channel consists of a pair of atoms in a specific 
hyperfine spin state $|f,m_f \rangle$.  The  closed channel consists of a 
pair of atoms in different hyperfine states with a higher scattering threshold.
The weak coupling between the channels is provided by the hyperfine 
interaction.  Since different hyperfine states have different magnetic 
moments, a magnetic field can be used to vary the energy gap 
between the scattering thresholds and bring a bound state in the 
closed channel into resonance with the threshold of the open channel.
The resulting enhancement of the scattering of particles 
in the open channel is a Feshbach resonance.
The scattering lengths generally 
vary slowly with the magnetic field $B$. If there is a Feshbach resonance
at $B_{\rm res}$, the scattering length varies dramatically with the 
magnetic field in the vicinity of $B_{\rm res}$. 
If the Feshbach resonance is narrow, the scattering length 
near the resonance has the approximate form
%----------------------
\begin{eqnarray}
a(B) \approx 
a_{\rm bg} \left( 1 -  {\Delta_{\rm res} \over B - B_{\rm res}} \right) \,.
\label{Feshbach}
\end{eqnarray}
%----------------------
If $B$ is far below or far above $B_{\rm res}$, the scattering length $a(B)$ 
has the off-resonant value $a_{\rm bg}$.
The parameter $\Delta_{\rm res}$, which controls the width of the resonance,
is defined so that $a(B)$ vanishes when $B = B_{\rm res}+ \Delta_{\rm res}$.
As $B$ increases through $B_{\rm res}$, $a(B)$ increases or decreases  
to $\pm \infty$, jumps discontinuously to $\mp \infty$, and then 
returns to its off-resonant value.
The magnetic field provides an experimental 
fine-tuning parameter that can be used to make $|a|$ arbitrarily large.
The existence of Feshbach resonances in atomic physics was 
first predicted in Ref.~\cite{TVS93} for the specific case of Cs atoms.
The use of a Feshbach resonance to produce a large scattering length 
in alkali atoms was first demonstrated by the MIT group using experiments 
with Bose-Einstein condensates of $^{23}$Na atoms \cite{Inouye98,Stenger99}
and by the Texas and JILA groups using experiments with ultracold gases 
of $^{85}$Rb atoms \cite{Courteille98,Roberts98}.  

For $^{23}$Na atoms, the spin-singlet and spin-triplet scattering 
lengths are both smaller than the natural scale $\ell_{\rm vdw}$,
so all the hyperfine spin states $|f, m_f\rangle$ 
have natural scattering lengths at $B=0$.  
However, there are Feshbach resonances 
at which the scattering lengths diverge.
For example, the $| 1,+1 \rangle$ hyperfine state has Feshbach resonances 
near 853 G and 907 G \cite{Inouye98} and the $| 1,-1 \rangle$ hyperfine state 
has a Feshbach resonance near 1195 G \cite{Stenger99}.
The Feshbach resonances near 853 G and 907 G were first observed
by the MIT group using a Bose-Einstein condensate of 
$^{23}$Na atoms \cite{Inouye98}.  They demonstrated that by varying 
the magnetic field they could change 
the scattering length by more than an order of magnitude.  
They also observed enhanced inelastic losses near the resonances.
In Ref.~\cite{Stenger99}, the inelastic losses were studied 
in greater detail.

For $^{85}$Rb atoms, the spin-singlet scattering length is large
and the spin-triplet scattering length is relatively large. 
They differ from the natural scale $\ell_{\rm vdw}$
by factors of about 17 and $-2.4$, respectively. 
Thus most of the hyperfine spin states have large scattering 
lengths at $B=0$.  
However, there are Feshbach resonances at which the scattering lengths 
diverge.  For example, the $|2,-2\rangle$ hyperfine state has a 
Feshbach resonance near 155 G.  This Feshbach resonance 
was first observed by the Texas group 
using an ultracold gas of $^{85}$Rb atoms~\cite{Courteille98}.  
The position and width of this Feshbach resonance 
were determined precisely by the JILA group~\cite{Roberts98}.  
They were used to improve the parameters 
of the atomic potential for Rb and to predict the scattering lengths 
of the hyperfine states of $^{85}$Rb and $^{87}$Rb.
They also demonstrated that by varying the magnetic field,
they could change the collision rate by 4 orders of magnitude 
and change the sign of the scattering length.

%%%%%%%%%%%%%%%%%%%%%%%%%%%%%%%%%%%%%%%%%%%%%%%%%%
%    The resonant and scaling limits
%%%%%%%%%%%%%%%%%%%%%%%%%%%%%%%%%%%%%%%%%%%%%%%%%%

\subsection{The resonant and scaling limits}
        \label{sec:scaling}
 
We have defined a large scattering length to be one that satisfies 
$|a| \gg \ell$, where $\ell$ is the natural low-energy length scale. 
The corrections to the universal behavior 
are suppressed by powers of $\ell/|a|$.
There are two obvious limits in which the size 
of these corrections decreases to zero:
\begin{itemize}

\item the {\it resonant} or {\it unitary limit}: 
$a \to \pm \infty$ with $\ell$ fixed,

\item the {\it scaling} or {\it zero-range limit}: 
$\ell \to 0$ with $a$ fixed.

\end{itemize}
It will sometimes also be useful to consider systems in which
the resonant and scaling limits are achieved simultaneously: 
$a = \pm \infty$ and $\ell = 0$.

The {\it resonant limit} is also sometimes called the {\it unitary limit},
because in this limit the S-wave contribution to the cross section 
at low energy saturates its unitarity bound $\sigma^{(L=0)} \le 8 \pi/k^2$.
The resonant limit can be approached by tuning 
the depth of the interatomic potential to a
critical value for which there is a 2-body bound state exactly 
at the 2-body threshold.
The resonant limit can also be approached 
by tuning the magnetic field to a Feshbach resonance. 
Since $a = \pm \infty$ in the {\it resonant limit}, 
one might expect that in this limit the natural low-energy length 
scale $\ell$ is the only important length scale at low energies.
This is true in the 2-body sector.
However, the Efimov effect reveals that there is another length scale
in the 3-body sector.
In the resonant limit, there are infinitely many, arbitrarily-shallow
3-body bound states with a spectrum of the form 
%----------------------
\begin{eqnarray}
E_T^{(n)} & \longrightarrow &
\left( e^{-2 \pi/s_0} \right)^{n-n_*} \hbar^2 \kappa_*^2/m\,,
%\nonumber \\ && 
\qquad {\rm as \ } n \to + \infty {\rm \ \ with \ } a = \pm \infty,
\label{B3-resonant}
\end{eqnarray}
%----------------------
%[E,H]
where $e^{-2 \pi/s_0} \approx 1/515.03$ for identical bosons and  
$\kappa_*$ can be interpreted as the approximate binding wave number
of the Efimov state labelled by the integer $n_*$.  
If we chose a different integer $n_*$, the  value of $\kappa_*$ 
would change by some power of $e^{\pi/s_0}$. 
Thus $\kappa_*$ is defined by Eq.~(\ref{B3-resonant})
only up multiplicative factors of $e^{\pi/s_0}$.

The {\it scaling limit} is also sometimes called the {\it zero-range limit},
because it can be reached by simultaneously tuning the range of the 
potential to zero and its depth to $\infty$ in such a way that the 
scattering length is fixed. 
The terminology  ``scaling limit'' seems to have been first used 
in Ref.~\cite{FTDA99}.
The scaling limit may at first seem a little contrived,
but it has proved to be a powerful concept.
It can be defined by specifying the phase shifts for 2-body scattering.
In the scaling limit, the S-wave phase shift $\delta_0(k)$ has
the simple form 
%----------------------
\begin{eqnarray}
k \cot \delta_0(k) = -1/a \,,
\label{kcot-scaling}
\end{eqnarray}
%----------------------
%[E,H]
and the phase shifts $\delta_L(k)$ for all higher partial waves vanish.
In the scaling limit, the scattering length $a$ sets the scale 
for most low-energy observables.
It is the only length scale in the 2-body sector.
However, as we shall see, in the 3-body sector, observables can also have 
logarithmic dependence on a second scale.  In the scaling limit,
there are infinitely many arbitrarily-deep
3-body bound states with a spectrum of the form \cite{MiF62,MiF62b}
%----------------------
\begin{eqnarray}
E_T^{(n)} & \longrightarrow &
\left( e^{-2 \pi/s_0} \right)^{n-n_*} {\hbar^2 \kappa_*^2 \over m}\,,
%\nonumber \\ && 
\qquad {\rm as \ } n \to - \infty {\rm \ \ with \ } \ell = 0 \,.
\label{B3-scaling}
\end{eqnarray}
%----------------------
%[E,H]
Thus the spectrum is characterized by
a parameter $\kappa_*$ with dimensions of wave number.

The scaling limit may appear to be pathological, because
the spectrum of 3-body bound states in Eq.~(\ref{B3-scaling})
is unbounded from below.
However, the deep 3-body bound states have a negligible effect 
on the low-energy physics of interest.
The pathologies of the scaling limit
can be avoided simply by keeping in mind 
that the original physical problem
before taking the scaling limit had a natural low-energy length scale $\ell$.
Associated with this length scale is a {\it natural energy scale} 
$\hbar^2 /m \ell ^2$.
Any predictions involving energies comparable to or larger than 
the natural low-energy length scale are artifacts of the scaling limit.
Thus when we use the scaling limit to describe a physical system, 
any predictions involving energies $|E| \gsim \hbar^2 /m \ell ^2$ 
should be ignored.  

In spite of its pathologies, we shall take the scaling limit 
as a starting point for describing atoms with large scattering length.
We will treat the deviations from the scaling limit as perturbations.
Our motivation is that when the scattering length is large,
there are intricate 
correlations between 3-body observables associated with the Efimov effect 
that can be easily lost by numerical approximations.
By taking the scaling limit, we can build in these intricate correlations
exactly at high energy.  Although these correlations are unphysical 
at high energy, this does not prevent us from describing
low-energy physics accurately.  It does, however, guarantee that the 
intricate 3-body correlations are recovered automatically
in the resonant limit $a \to \pm \infty$.

%%%%%%%%%%%%%%%%%%%%%%%%%%%%%%%%%%%%%%%%%%%%%%%%%%
%    Hyperspherical coordinates
%%%%%%%%%%%%%%%%%%%%%%%%%%%%%%%%%%%%%%%%%%%%%%%%%%

\subsection{Hyperspherical coordinates}
        \label{subsec:hyper}

The universal aspects of the 3-body problem can be understood most easily by
formulating it in terms of {\it hyperspherical coordinates}.
A good introduction to hyperspherical coordinates
and a thorough review of the hyperspherical
formalism is given in a recent review 
by Nielsen, Fedorov, Jensen, and Garrido \cite{NFJG01}.

In order to define hyperspherical coordinates, 
we first introduce Jacobi coordinates.  A set of Jacobi coordinates 
consists of the separation vector ${\bm r}_{ij}$ between a pair of
atoms and the separation vector ${\bm r}_{k,ij}$ of the third atom from the
center-of-mass of the pair. For atoms of equal mass, the Jacobi 
coordinates are
%----------------------
%\begin{subequations}
\begin{eqnarray}
{\bm r}_{ij} & = & {\bm r}_i - {\bm r}_j \,, 
\qquad
{\bm r}_{k, ij}  =  {\bm r}_k - \mbox{$1\over2$}({\bm r}_i + {\bm r}_j) \,.
\label{jacobi}
\end{eqnarray}
%----------------------
%[E,H]
The {\it hyperradius} $R$ is the root-mean-square 
separation of the three atoms:
%----------------------
%\begin{subequations}
\begin{eqnarray}
R^2 &=& \mbox{$1 \over 3$} \left( r_{12}^2 + r_{23}^2 + r_{31}^2 \right) 
= \mbox{$1 \over 2$} r_{ij}^2 + \mbox{$2 \over 3$} r_{k,ij}^2 \,.
\label{hyperradius}
\end{eqnarray}
%\end{subequations}
%----------------------
%[H,E]
The hyperradius is small only if all three atoms are close together. 
It is large if any single atom is far from the other two. 
The {\it Delves hyperangle} \cite{Delves60} $\alpha_k$ is defined by
%----------------------
\begin{eqnarray}
\alpha_k=\arctan \left( {\sqrt{3} r_{ij} \over 2 r_{k,ij}} \right) \,,
\label{alphak}
\end{eqnarray}
%----------------------
%[H,E]
where $(i,j,k)$ is a permutation of $(1,2,3)$. The range of the hyperangle
$\alpha_k$ is from 0 to ${1\over2}\pi$.  It is near 0 when atom $k$ 
is far from atoms $i$ and $j$, and it is near ${1\over2}\pi$ 
when atom $k$ is near the center of mass of atoms $i$ and $j$.

The Schr{\"o}dinger equation for the stationary wave function
$\Psi({\bm r}_1, {\bm r}_2, {\bm r}_3)$ of three atoms with mass $m$
interacting through a potential $V$ is
%----------------------
\begin{eqnarray}
\left(-{\hbar^2 \over 2m} \sum_{i=1}^3 \nabla_i^2 + V ({\bm r}_1, {\bm r}_2,
{\bm r}_3)\right) \Psi = E \Psi \,.
\label{Seq:1}
\end{eqnarray}
%----------------------
%[H,E] 
The wave function $\Psi$ in the center-of-mass frame depends on 
6 independent coordinates.  A convenient choice consists of 
the hyperradius $R$, one of the hyperangles $\alpha_k$, and the
unit vectors $\hat {\bm r}_{ij}$ and $\hat {\bm r}_{k,ij}$.
We will refer to the 5 dimensionless variables 
($\alpha_k,\hat {\bm r}_{ij},\hat {\bm r}_{k,ij}$) as hyperangular
variables and denote them collectively by $\Omega$.
When expressed in terms of hyperspherical coordinates, the
Schr{\"o}dinger equation reduces to
%----------------------
\begin{eqnarray}
\left[ T_R + T_\Omega + V (R, \Omega) \right] \Psi = E \Psi \,,
\label{Seq:3}
\end{eqnarray}
%----------------------
%
%[h!,e!]
where $T_R$ is the hyperradial kinetic energy operator,
%----------------------
%\begin{subequations}
\begin{eqnarray}
T_R & = & - {\hbar^2 \over 2m}
\left[ {\partial^2 \ \over \partial R^2}
     + {5 \over R} {\partial \ \over \partial R} \right]
%\\& = & 
={\hbar^2 \over 2m} R^{-5/2}
\left [ - {\partial^2 \ \over \partial R^2} 
     + {15 \over 4 R^2} \right] R^{5/2} \,,
\end{eqnarray}
%\end{subequations}
%----------------------
%[E]
and $T_\Omega$ is the kinetic energy operator associated 
with the hyperangular variables.

The {\it Faddeev equations} are a set of three differential equations 
equivalent to the 3-body Schr{\"o}dinger equation that exploit 
the simplifications associated with configurations consisting 
of a 2-body cluster that is well-separated from the third atom.
The solutions to the Faddeev equations are three Faddeev 
wavefunctions whose sum is a solution to the Schr{\"o}dinger equation:
%----------------------
\begin{eqnarray}
\Psi ({\bm r}_1, {\bm r}_2, {\bm r}_3) 
&=& \psi^{(1)}({\bm r}_{23},{\bm r}_{1,23}) 
+ \psi^{(2)}({\bm r}_{31},{\bm r}_{2,31}) 
%\nonumber \\ && 
+ \psi^{(3)}({\bm r}_{12},{\bm r}_{3,12}) \,.
\label{Psi-Faddev}
\end{eqnarray}
%----------------------
%[E,H]
We restrict our attention to states with 
total angular momentum quantum number $L=0$.
At low energies, we can make an additional simplifying assumption 
of neglecting subsystem angular momenta.
If the three particles are identical bosons, the  
three Faddeev wave functions can be expressed in terms of a
single function $\psi(R, \alpha)$:
%----------------------
\begin{eqnarray}
\Psi ({\bm r}_1, {\bm r}_2, {\bm r}_3) 
= \psi(R, \alpha_1) + \psi(R,\alpha_2) + \psi(R, \alpha_3) \,.
\label{psi3}
\end{eqnarray}
%----------------------
%[H,E]
The three Faddeev equations can be reduced to a single 
integro-differential equation for the Faddeev wavefunction
$\psi(R, \alpha)$.
(See Ref.~\cite{bigrev} for more details.)

A convenient way to solve the resulting equation is to use a
{\it hyperspherical expansion}.
For each value of $R$, the wave function $\psi (R,\alpha)$
is expanded in a complete set of functions $\phi_n(R, \alpha)$ of the
hyperangle $\alpha$:
%----------------------
\begin{eqnarray}
\psi(R, \alpha) = {1 \over R^{5/2} \sin(2 \alpha)} 
                        \sum_n f_n (R) \phi_n(R, \alpha) \,.
\label{Faddeev-exp}
\end{eqnarray}
%----------------------
%[H,E]
The functions $\phi_n(R, \alpha)$ are solutions to an
eigenvalue equation in the hyperangle $\alpha$ that is parametric in
the hyperradius $R$.  
The eigenvalues $\lambda_n(R)$ 
determine channel potentials for the hyperradial variable:
%----------------------
\begin{eqnarray}
\label{Vn}
V_n(R) = [\lambda_n (R) -4] {\hbar^2 \over 2m R^2} \,.
\label{Vch}
\end{eqnarray}
%----------------------
%[H,E]
The hyperradial wavefunctions $f_n (R)$ satisfy an infinite 
set of coupled partial differential equations.
In the {\it adiabatic hyperspherical approximation} \cite{ZM88}, 
the coupling terms are neglected and the equations decouple.
They reduce to independent hyperradial equations
for each of the hyperspherical potentials:
%----------------------
\begin{eqnarray}
\left[ {\hbar^2 \over 2 m} 
\left( - {\partial^2 \ \over \partial R^2} + {15\over 4 R^2} \right) 
        + V_n(R)\right] f_n(R)
\approx E f_n (R) \,.
%\nonumber\\
\label{aha:F}
\end{eqnarray}
%----------------------
%[E]
In the {\it hyperspherical close-coupling approximation} \cite{ZM88},
the diagonal coupling terms are also retained.  This approximation 
is more accurate, because it is variational in character.

%\newpage

%%%%%%%%%%%%%%%%%%%%%%%%%%%%%%%%%%%%%%%%%%%%%%%%%%
%
%     UNIVERSALITY FOR TWO IDENTICAL BOSONS
%
%%%%%%%%%%%%%%%%%%%%%%%%%%%%%%%%%%%%%%%%%%%%%%%%%%

\section{Two Identical Bosons}
        \label{sec:uni2}

%%%%%%%%%%%%%%%%%%%%%%%%%%%%%%%%%%%%%%%%%%%%%%%%%%
%    Atom-atom scattering
%%%%%%%%%%%%%%%%%%%%%%%%%%%%%%%%%%%%%%%%%%%%%%%%%%

In this section, we summarize the universal properties of two 
identical bosons in the scaling limit in which the 
large scattering length is the only length scale.

\subsection{Atom-atom scattering}
\label{sec:uni2AA}

The cross section for low-energy atom-atom scattering
is a universal function of the scattering length 
and the collision energy $E$. 
By low energy, we mean $E=\hbar^2k^2/m$ 
much smaller than the natural low-energy scale $\hbar^2/m \ell^2$. 
The partial wave expansion in Eq.~(\ref{pwe})
expresses the scattering amplitude in terms of phase shifts
$\delta_L (k)$. 
In the scaling limit, all the phase shifts vanish except for
$\delta_0(k)$ and all the coefficients in the
low-energy expansion of $k\cot\delta_0(k)$ vanish with the 
exception of the leading term $-1/a$. 
The scattering amplitude in Eq.~(\ref{pwe}) reduces to
%----------------------
\begin{eqnarray}
f_k(\theta) = {1 \over - 1/a -ik} \,,
\label{f-2}
\end{eqnarray}
%----------------------
%[H,E] 
and the differential cross section in Eq.~(\ref{dcross}) is
%----------------------
\begin{eqnarray}
{d\sigma_{AA} \over d\Omega} = {4a^2 \over 1 + a^2 k^2} \,.
\label{sig-2}
\end{eqnarray}
%----------------------
%[H,E]
The cross section is obtained by integrating 
over the solid angle $2 \pi$.
For wave numbers $k \gg 1/|a|$, the differential cross section has the
scale-invariant form $4/k^2$, which saturates the upper bound from
partial-wave unitarity in the $L=0$ channel.

The wave function for atom-atom scattering states at
long distances is a universal function of the scattering length 
and the separation $r$. By long distances, we mean 
$r$ much larger than the natural low-energy length scale $\ell$.  
The stationary wave function 
in the center-of-mass frame for two atoms in an $L=0$ state
with energy $E = \hbar^2k^2/m$ is
%----------------------
\begin{eqnarray}
\psi_{AA}({\bm r})= {1 \over r} \sin \left[ kr - \arctan(ka) \right] \,.
\label{psi-AAsin}
\end{eqnarray}
%----------------------
%[e!,h!]
This wave function satisfies the boundary condition
%----------------------
\begin{eqnarray}
\psi_{AA}({\bm r}) \longrightarrow C
\left({1 \over r} - {1 \over a}\right)\,,
\hspace{1cm} \mbox{as }  r \to 0 \,,
\label{psi-AAbc}
\end{eqnarray}
%----------------------
%[e!,h!]
where $C$ is an arbitrary normalization constant.
In the scaling limit $\ell \to 0$,
the effects of short-distances $r \sim \ell$
enter only through this boundary condition.

%%%%%%%%%%%%%%%%%%%%%%%%%%%%%%%%%%%%%%%%%%%%%%%%%%
%    The shallow dimer
%%%%%%%%%%%%%%%%%%%%%%%%%%%%%%%%%%%%%%%%%%%%%%%%%%

\subsection{The shallow dimer}
\label{sec:uni2D}

The spectrum of shallow 2-body bound states is also universal.  
By a {\it shallow bound state},
we mean one whose binding energy $E_D$ is
much smaller than the natural low-energy scale $\hbar^2/m \ell^2$.
For $a<0$, there are no shallow bound states.  
For $a>0$, there is a single shallow bound state, 
which we will refer to as the {\it shallow dimer},
or simply as the {\it dimer} for brevity. 
The binding energy $E_D$ of the dimer in the scaling limit is
%----------------------
\begin{eqnarray}
E_D={\hbar^2 \over ma^2}  \,.
\label {B2-uni}
\end{eqnarray}
%----------------------
%[H,E]

The wave function of the dimer at separations $r \gg \ell$ 
is also universal:
%----------------------
\begin{eqnarray}
\psi_D({\bm r})=\frac{1}{r} e^{-r/a} \,.
\label{psi-D}
\end{eqnarray}
%----------------------
%[H,E]
The size of the dimer is roughly $a$. 
A quantitative measure of the size is the 
mean separation of the atoms:  $\langle r \rangle = a/2$.
The wavefunction in Eq.~(\ref{psi-D}) satisfies the boundary condition in 
Eq.~(\ref{psi-AAbc}).

%%%%%%%%%%%%%%%%%%%%%%%%%%%%%%%%%%%%%%%%%%%%%%%%%%
%    Continuous scaling symmetry
%%%%%%%%%%%%%%%%%%%%%%%%%%%%%%%%%%%%%%%%%%%%%%%%%%

\subsection{Continuous scaling symmetry}

The only parameter in the universal expressions for the cross-section 
in Eq.~(\ref{sig-2}) and the binding energy in Eq.~(\ref{B2-uni}) 
is the scattering length $a$. 
The fact that low-energy observables depend only on a single 
dimensionful parameter $a$ can be expressed 
formally in terms of a {\it continuous scaling symmetry} that
consists of rescaling $a$, 
the coordinate ${\bm r}$, and the time $t$ by appropriate powers 
of a positive number $\lambda$:
%----------------------
\begin{eqnarray}
a & \longrightarrow & \lambda a \,,
\qquad
{\bm r} \longrightarrow  \lambda {\bm r} \,,
\qquad
t  \longrightarrow  \lambda^2 t \,.
\label{csi}
\end{eqnarray}
%----------------------
%[E,H]
The scaling of the time by the square of the scaling factor 
for lengths is natural in a nonrelativistic system.
Under this symmetry, observables, such as the dimer binding energy 
$E_D$ or the atom-atom cross section $\sigma_{AA}$, 
scale with the powers of $\lambda$ suggested by dimensional analysis.

The scaling symmetry strongly constrains the dependence of the
observables on the scattering length and on kinematic variables.
As a simple example, the dimer binding energy scales as 
$E_D \to \lambda^{-2} E_D$.  The scaling symmetry constrains 
its dependence on the scattering length to be proportional to $1/a^2$, 
in agreement with the explicit formula in Eq.~(\ref{B2-uni}).
As another example, the atom-atom cross section
scales as $\sigma_{AA} \to \lambda^{2} \sigma_{AA}$.
The scaling symmetry constrains its dependence
on the energy and the scattering length:
%----------------------
\begin{eqnarray}
\sigma_{AA}(\lambda^{-2}E; \lambda  a) = \lambda^{2} \sigma_{AA}(E; a) \,.
\end{eqnarray}
%----------------------
%[H,E]
The explicit expression for the differential cross section in
Fig.~(\ref{sig-2}) is consistent with this constraint.

The scattering length $a$ changes discontinuously between $+\infty$ 
and $-\infty$ as the system is tuned through its critical point.
Since $1/a$ changes smoothly, this is a more convenient interaction 
variable.  To exhibit the scaling symmetry most clearly, 
it is convenient to use an energy variable with the same dimensions 
as the interaction variable.  
A convenient choice is the wave number variable
%----------------------
\begin{eqnarray}
K = {\rm sign} (E) (m|E|/\hbar^2)^{1/2} \,.
\label{K-def}
\end{eqnarray}
%----------------------
%[H,E]
The set of all possible low-energy 2-body states in the scaling limit 
can be represented as points $(a^{-1}, K)$ on the plane 
whose horizontal axis is $1/a$ and whose vertical axis is $K$.
It is convenient to also introduce polar coordinates consisting of
a radial variable $H$ and an angular variable $\xi$ defined by
%----------------------
\begin{eqnarray}
1/a &=& H \cos \xi \,,
\qquad
K   = H \sin \xi \,.
\label{Hxi-def}
\end{eqnarray}
%----------------------
%[E,H]
In terms of these polar coordinates,
the scaling symmetry given by Eqs.~(\ref{csi}) is simply a rescaling 
of the radial variable: $H \to \lambda ^{-1} H$.

%%%%%%%%%%%%%%%%%%%%%%%%%%%%%%%%%%%%%%%%%%%%%%%%%%
\begin{figure}[htb]
\bigskip
\centerline{\includegraphics*[width=8cm,angle=0]{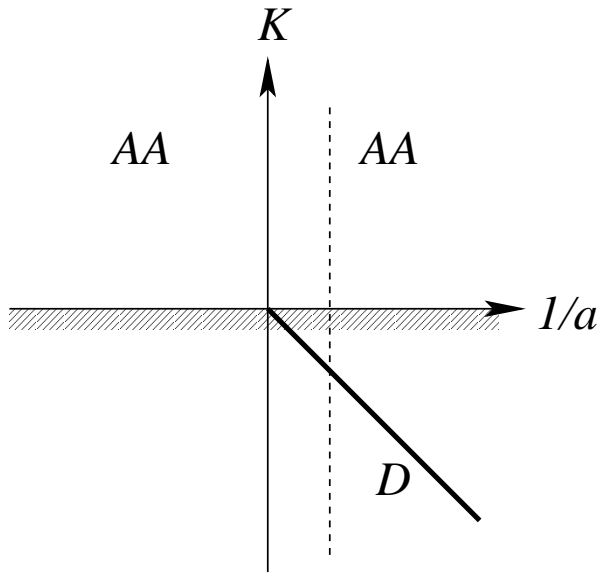}}
\medskip
\caption
{The $a^{-1}$--$K$ plane for the 2-body problem. The allowed region for
atom-atom scattering states are the two quadrants labelled $AA$. 
The heavy line labelled $D$ is the shallow dimer.
The cross-hatching indicates the 2-atom threshold.}
\label{fig:2body}
\end{figure}
%%%%%%%%%%%%%%%%%%%%%%%%%%%%%%%%%%%%%%%%%%%%%%%%%%

The $a^{-1}$--$K$ plane for the 2-body system in the scaling limit
is shown in Fig.~\ref{fig:2body}.
The possible states are atom-atom scattering states ($AA$)
and the shallow dimer ($D$). 
The threshold for atom-atom scattering states is indicated
by the hatched area.
The shallow dimer is represented by the heavy line given
by the ray $\xi = -{1\over 4}\pi$.
A given physical system has a specific value of the scattering length, 
and so is represented by a vertical line, 
such as the dashed line in Fig.~\ref{fig:2body}.
Changing $a$ corresponds to sweeping the line horizontally across the page. 
The resonant limit corresponds to tuning the vertical line to the $K$ axis.

%%%%%%%%%%%%%%%%%%%%%%%%%%%%%%%%%%%%%%%%%%%%%%%%%%
%    Scaling violations
%%%%%%%%%%%%%%%%%%%%%%%%%%%%%%%%%%%%%%%%%%%%%%%%%%

\subsection{Scaling violations}
\label{sec:2bodyscalviol}

The continuous scaling symmetry is a trivial consequence of the fact 
that $a$ is the only length scale that remains nonzero in the scaling limit.
For real atoms, the scaling limit can only be an approximation.
There are {\it scaling violations} that give corrections 
that are suppressed by powers of $\ell/|a|$.
In the 2-body sector, the most important scaling violations 
come from the S-wave effective range $r_s$
defined by the effective-range expansion in Eq.~(\ref{kcot}). 

The differential cross section for atom-atom scattering
can be expanded in powers of $\ell/a$ with $a k$ fixed:
%----------------------
\begin{eqnarray}
{d \sigma \over d \Omega} &=& 
{4 a^2 \over 1 + a^2 k^2}
\left( 1 + {r_s \over a} \, {a^2 k^2 \over 1 + a^2 k^2} 
        + \ldots \right) \,.
\label{dsig:exp}
\end{eqnarray}
%----------------------
%[H,E]
The leading term is the universal expression in Eq.~(\ref{sig-2}).
The next-to-leading term is determined by $r_s$. 

If $a>0$, we can also consider the scaling violations to the 
binding energy of the shallow dimer. The binding energy
can be expanded in powers of $\ell/a$:
%----------------------
\begin{eqnarray}
E_2^{(-)} \approx {\hbar^2 \over m a^2} \left ( 1 + {r_s \over a} + {5 r_s^2
\over 4 a^2} + \ldots \right ) \,.
\label{B2-exp}
\end{eqnarray}
%----------------------
%[H,E]
The leading term is the universal expression in Eq.~(\ref{B2-uni}).
The first two correction terms are determined by $r_s$.

%\newpage

%%%%%%%%%%%%%%%%%%%%%%%%%%%%%%%%%%%%%%%%%%%%%%%%%%
%
%     UNIVERSALITY FOR THREE IDENTICAL BOSONS
%
%%%%%%%%%%%%%%%%%%%%%%%%%%%%%%%%%%%%%%%%%%%%%%%%%%

\section{Three Identical Bosons}
        \label{sec:uni3}

In this section, we summarize the universal properties of three
identical bosons in the scaling limit in which the 
only scales are those provided by the large scattering length $a$
and the Efimov parameter $\kappa_*$.

%%%%%%%%%%%%%%%%%%%%%%%%%%%%%%%%%%%%%%%%%%%%%%%%%%
%    Boundary condition at short distances
%%%%%%%%%%%%%%%%%%%%%%%%%%%%%%%%%%%%%%%%%%%%%%%%%%

\subsection {Boundary condition at short distances}
        \label{subsec:bc}

When the scattering length is large compared to the natural 
low-energy length scale $\ell$, the range of the hyperradius $R$ 
includes four important regions.
It is useful to give names to each of these regions:
\begin{itemize}
\item the {\it short-distance region} $R \lsim |\ell|$,
\item the {\it scale-invariant region} $\ell \ll R \ll |a|$,
\item the {\it long-distance region} $R \sim |a|$,
\item the {\it asymptotic region} $R \gg |a|$.
\end{itemize}
In the scaling limit $\ell \to 0$, the short-distance region
shrinks to 0.  Its effects can however be taken into account 
through a boundary condition on the hyperradial wavefunction 
in the scale-invariant region.

\begin{figure}[htb]
\bigskip
\centerline{\includegraphics*[width=9cm,angle=0]{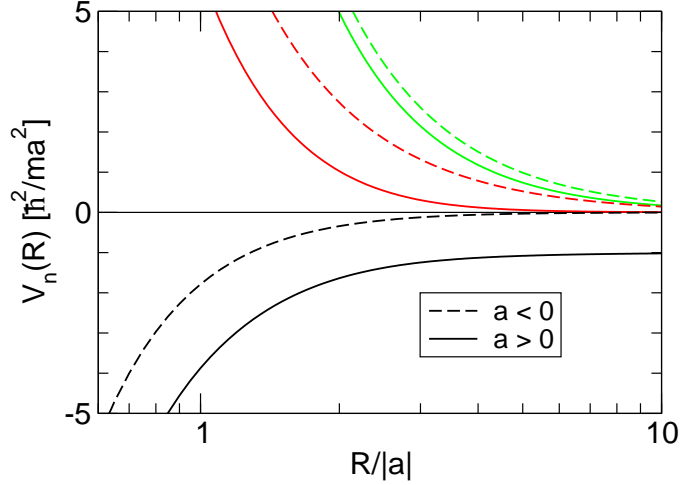}}
\medskip
\caption{The three lowest hyperspherical potentials $V_n(R)$ 
scaled by $\hbar^2/ma^2$
for $a > 0$ (solid lines) and for $a < 0$ (dashed lines).}
\label{fig:hap}
\end{figure}

The hyperspherical expansion of the Faddeev wave function 
is given in Eq.~(\ref{Faddeev-exp}).
In the scaling limit, the channel eigenvalues $\lambda_n(R)$ 
determined by the eigenvalue equation 
for the hyperangular functions $\phi_n(R,\alpha)$
satisfy \cite{Efimov71}
%----------------------
\begin{eqnarray}
\cos \left ( \lambda^{1/2} \mbox {$\pi \over 2$} \right)
- {8 \over \sqrt {3}} \lambda^{-1/2} 
        \sin \left ( \lambda^{1/2} \mbox {$\pi \over 6$} \right )
%\nonumber  \\ 
= \sqrt{2} \lambda^{-1/2} \sin \left ( \lambda^{1/2} \mbox {$\pi \over 2$} 
  \right ) {R \over a} \,.
\label{cheigen}
\end{eqnarray}
%----------------------
%[H,E]
The lowest channel potentials $V_n(R)$ defined by Eq.~(\ref{Vch})
are shown in Fig.~\ref{fig:hap}.
%%%%%%%%%%%%%%%%%%%%%%%%%%%%%%%%%%%%%%%%%%%%%%%%%%
%%%%%%%%%%%%%%%%%%%%%%%%%%%%%%%%%%%%%%%%%%%%%%%%%%
For $R \ll |a|$, the channel eigenvalues approach constants $\lambda_n(0)$
and the channel potentials in Eq.~(\ref{Vch}) 
have the scale-invariant behavior $1/R^2$.
Whether $a$ is positive or negative, the lowest eigenvalue 
$\lambda_0(R)$ is negative at $R=0$.
It can be expressed as $\lambda_0(0)= - s_0^2$, 
where $s_0 \approx 1.00624$ is the solution to the transcendental equation 
%----------------------
\begin{eqnarray}
s_0 \cosh {\pi s_0 \over 2} = {8 \over \sqrt{3}} \sinh {\pi s_0 \over 6} \,.
\label{s0}
\end{eqnarray}
%----------------------
%[H,E]
Thus the potential $V_0(R)$ is an attractive $1/R^2$ potential
for $R \ll |a|$:
%----------------------
\begin{eqnarray}
V_0 (R) \approx - (4 + s_0^2) {\hbar^2 \over 2mR^2}, 
\hspace{1cm} \ell \ll  R \ll |a| \,.
\label{V0-si}
\end{eqnarray}
%----------------------
%[H,E]
All the channel eigenvalues for $n \ge 1$ are positive at $R=0$:
$\lambda_n(0)\ge 19.94$.
Thus all the channel potentials $V_n(R)$ for $n \ge 1$ are repulsive 
$1/R^2$ potentials in the region $R \ll |a|$.  
The hyperradial wave functions $f_n(R)$
for $n \ge 1$ therefore decrease exponentially for $R \ll |a|$.

In the scaling limit $\ell \to 0$, the lowest adiabatic hypersherical 
potential in Eq.~(\ref{V0-si}) behaves like $1/R^2$ 
all the way down to $R=0$.  Such a potential is too singular 
to have well-behaved solutions.
If the exact solution $f (R)$ for the hyperradial wave function
at short distances was known, 
it could be matched onto the solution in the scaling limit
by choosing a hyperradius $R_0$ in the scale-invariant region and
demanding that the logarithmic derivatives $R_0 f^\prime (R_0) / f (R_0)$ 
match at that point.  
If we also choose $R_0 \ll (m|E|/\hbar^2)^{-1/2}$, the energy eigenvalue $E$
in Eq.~(\ref{aha:F}) can be neglected relative to the channel potential.
The most general solution for the hyperradial wave function 
in the scale-invariant region is 
%----------------------
\begin{eqnarray}
f(R) = R^{1/2} \left[A e^{is_0 \ln (H R)} + B e^{-is_0 \ln (H R)}
\right] \,,
\label{f-general}
\end{eqnarray}
%----------------------
%[H,E]
where $A$ and $B$ are arbitrary coefficients.  
To make the arguments of the logarithms dimensionless,
we have inserted factors of $H$, the wavenumber variable defined in
Eq.~(\ref{Hxi-def}).  The terms with the coefficients $A$ and $B$ represent 
an outgoing hyperradial wave and an incoming hyperradial wave, 
respectively.
If $|A|<|B|$,
there is a net flow of probability into the short-distance region.
As will be discussed in detail in Section~\ref{sec:deep},
such a flow of probability is possible if there are 
deep diatomic molecules. 
In this section, we assume that there are no deep 2-body bound states.
The probability in the incoming hyperradial wave must 
therefore be totally reflected at short distances. 
This requires $|A| = |B|$, which implies that $A$ and $B$ 
differ only by a phase:
%----------------------
\begin{eqnarray}
A = - e^{2 i\theta_*} B
\label{AB}
\end{eqnarray}
%----------------------
%[E]
for some angle $\theta_*$.  
This angle can be expressed as $\theta_*= - s_0 \ln (H / \Lambda_0)$,
where $\Lambda_0$ is the product of $1/R_0$ and a complicated function 
of  $R_0 f'(R_0)/ f(R_0)$.  The effects of the short-distance
region on $f(R)$ enter only through the wavenumber 
variable $\Lambda_0$. By solving the hyperradial equation for the
scale-invariant potential in Eq.~(\ref{V0-si}), we find that
$\Lambda_0$ differs from the 3-body parameter 
$\kappa_*$ defined by the Efimov spectrum in the resonant limit
only by a multiplicative numerical constant.  Thus the angle $\theta_*$ 
in Eq.~(\ref{AB}) can be 
expressed as 
%----------------------
\begin{eqnarray}
\theta_* = - s_0 \ln (c H / \kappa_*) \,,
\label{theta-star}
\end{eqnarray}
%----------------------
%[H,E]
where $H$ is defined in Eq.~(\ref{Hxi-def}) and
$c$ is a constant.

%%%%%%%%%%%%%%%%%%%%%%%%%%%%%%%%%%%%%%%%%%%%%%%%%%
%    Discrete scaling symmetry
%%%%%%%%%%%%%%%%%%%%%%%%%%%%%%%%%%%%%%%%%%%%%%%%%%

\subsection{Discrete scaling symmetry}
        \label{sec:discrete}

The boundary condition in Eq.~(\ref{AB}) gives
{\it logarithmic scaling violations} that give corrections 
to the scaling limit that are functions of $\ln(H/\kappa_*)$.
Logarithmic scaling violations do not become less important 
as one approaches the scaling limit,
and therefore cannot be treated as perturbations.
Because the boundary condition enters through the phase in Eq.~(\ref{AB}), 
the logarithmic scaling violations must be log-periodic 
functions of $H/\kappa_*$ with period $\pi/s_0$.

The 3-body sector in the scaling limit has a trivial continuous 
scaling symmetry defined by Eqs.~(\ref{csi}) together with 
$\kappa_* \to \lambda^{-1} \kappa_*$. However, 
because of the log-periodic form of the logarithmic scaling violations,
it also has a nontrivial {\it discrete scaling symmetry}.
There is a discrete subgroup of the continuous scaling symmetry
that remains an exact symmetry in the scaling limit:  

%----------------------
\begin{eqnarray}
\kappa_* & \longrightarrow & \kappa_* \,,
\qquad
a \longrightarrow \lambda_0^n a \,,
\qquad
{\bm r} \longrightarrow \lambda_0^n {\bm r} \,,
\qquad
t \longrightarrow \lambda_0^{2n} t \,,
\label{dsi}
\end{eqnarray}
%----------------------
%[E,H]
where $n$ is an integer, $\lambda_0 = e^{\pi/s_0}$, 
and $s_0 \approx 1.00624$ is the solution to the transcendental equation 
in Eq.~(\ref{s0}).  The numerical value of the discrete scaling factor 
is $e^{\pi/s_0} \approx 22.7$.
Under the discrete scaling  symmetry, 3-body observables,
such as binding energies and cross sections,
scale with the integer powers of $\lambda_0$
suggested by dimensional analysis.
By combining the trivial continuous scaling symmetry
with the discrete scaling symmetry given by Eqs.~(\ref{dsi}), 
we can see that $\kappa_*$ is only defined modulo multiplicative factors 
of $\lambda_0$.

The discrete scaling symmetry strongly constrains the dependence 
of the observables on 
the parameters $a$ and $\kappa_*$ and on kinematic variables. 
For example,  the scaling of the atom-dimer cross section is 
$\sigma_{AD} \to (\lambda_0^{m})^2 \sigma_{AD}$.  
The discrete scaling symmetry constrains its dependence 
on $a$, $\kappa_*$, and the energy $E$: 
%----------------------
\begin{eqnarray}
\sigma_{AD} (\lambda_0^{-2m} E; \lambda_0^{m} a, \kappa_*)
= \lambda_0^{2m}  \sigma_{AD} (E; a, \kappa_*) \,,
\label{sigAD:dss}
\end{eqnarray}
%----------------------
%[H,E]
for all integers $m$.  At $E=0$, the cross section is simply 
$\sigma_{AD} = 4 \pi | a_{AD} |^2$,
where $a_{AD}$ is the atom-dimer scattering length.
The constraint in Eq.~(\ref{sigAD:dss}) implies that the 
atom-dimer scattering length $a_{AD}$ is proportional to $a$
with a coefficient that is a log-periodic function of $a  \kappa_*$
with period $\pi/s_0$.  The explicit expression for the atom-dimer
scattering length is given in Eq.~(\ref{a12-explicit})
and it is indeed consistent with this constraint.

To illustrate the discrete scaling symmetry, it is convenient 
to use the interaction variable $1/a$ and the energy variable $K$
defined in Eq.~(\ref{K-def}).
For a given value of $\kappa_*$,
the set of all possible low-energy 3-body states in the scaling limit 
can be represented as points $(a^{-1}, K)$ on the plane 
whose horizontal axis is $1/a$ 
and whose vertical axis is $K$.
The discrete scaling transformation in Eqs.~(\ref{dsi})
is simply a rescaling of the radial variable 
$H$ defined in Eq.~(\ref{Hxi-def}) with 
$\kappa_*$ and $\xi$ fixed:  $H \to \lambda_0^{-m} H$.

%%%%%%%%%%%%%%%%%%%%%%%%%%%%%%%%%%%%%%%%%%%%%%%%
\begin{figure}[htb]
\bigskip
\centerline{\includegraphics*[width=8cm,angle=0]{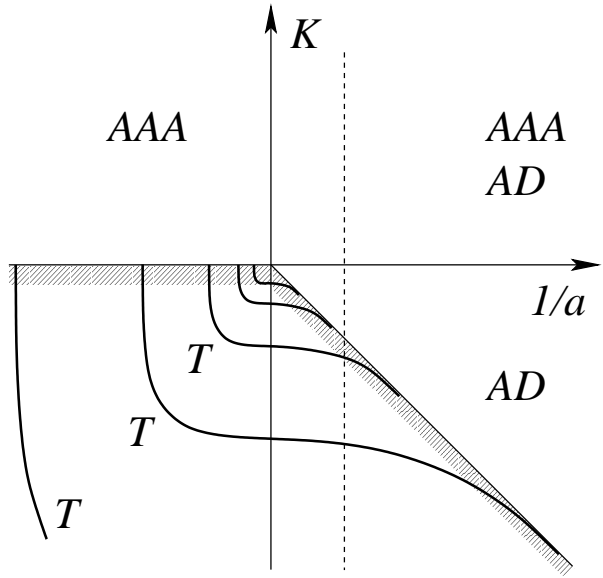}}
\medskip
\caption
{The $a^{-1}$--$K$ plane for the 3-body problem. The allowed regions for
3-atom scattering states and for atom-dimer scattering states are
labelled $AAA$ and $AD$, respectively. The heavy lines labeled $T$
are three of the infinitely many branches of Efimov states. 
The cross-hatching indicates the threshold for scattering states.
}
\label{fig:3body}
\end{figure}
%%%%%%%%%%%%%%%%%%%%%%%%%%%%%%%%%%%%%%%%%%%%%%%%%

The $a^{-1}$--$K$ plane for three identical bosons in the scaling limit 
is shown in Fig.~\ref{fig:3body}.
The possible states are 3-atom scattering states ($AAA$),
atom-dimer scattering states ($AD$), and Efimov trimers ($T$).
The threshold for scattering states is indicated by the hatched area.
The Efimov trimers are represented by the heavy
lines below the threshold.\footnote{The curves for the trimer 
binding energies in Fig.~\ref{fig:3body}
actually correspond to plotting $H^{1/4}\sin\xi$ versus $H^{1/4}\cos\xi$.
This effectively reduces the discrete scaling
factor 22.7 down to $22.7^{1/4} = 2.2$, allowing a greater range of
$a^{-1}$ and $K$ to be shown in the Figure.} 
There are infinitely many branches of Efimov trimers, 
but only a few are shown.
They intercept the vertical axis at the points
$K = - (e^{-\pi/s_0} )^{n-n_*} \kappa_*$.
A given physical system has a specific value of the scattering length, 
and so is represented by a vertical line.
The resonant limit corresponds to tuning the vertical line to the $K$ axis.

%%%%%%%%%%%%%%%%%%%%%%%%%%%%%%%%%%%%%%%%%%%%%%%%
\begin{figure}[htb]
\bigskip
\centerline{\includegraphics*[width=8cm,angle=0]{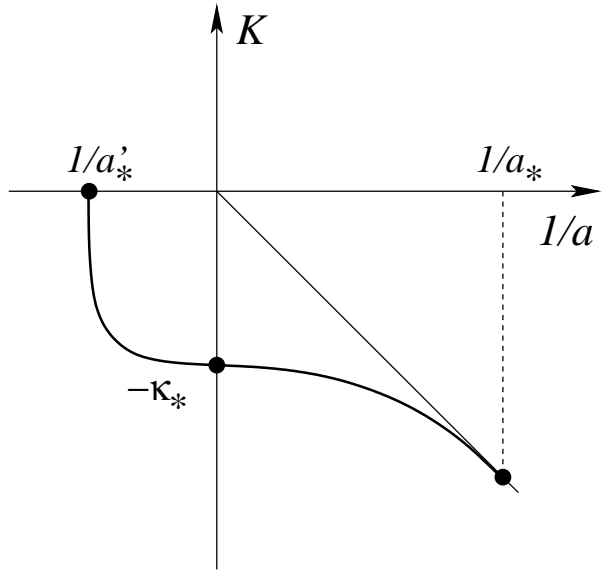}}
\medskip
\caption
{The energy variable $K$ for the branch of Efimov trimers 
labelled by $n=n_*$ as a function of $1/a$. 
In the resonant limit, the binding wave number is $\kappa_*$.
The branch disappears through the atom-dimer threshold at
$a=a_*$ and through the 3-atom threshold at $a = a_*'$.
}
\label{fig:efistar}
\end{figure}
%%%%%%%%%%%%%%%%%%%%%%%%%%%%%%%%%%%%%%%%%%%%%%%%%

Changing $1/a$ continuously from a large positive value
to a large negative value corresponds to sweeping the vertical dashed line in
Fig.~\ref{fig:3body} from right to left across the 
page.  The Efimov trimers appear one by one at the 
atom-dimer threshold at positive critical values of $1/a$ 
that differ by powers of $e^{\pi/s_0} \approx 22.7$ 
until there are infinitely many at $1/a = 0$.
As $1/a$ continues to decrease through negative values, the Efimov trimers
disappear one by one through the 3-atom threshold
at negative critical values of $1/a$ that differ by powers 
of $e^{\pi/s_0}$.  We will focus on the specific branch of Efimov trimers 
labelled by the integer $n=n_*$, 
which is illustrated in Fig.~\ref{fig:efistar}.
At some positive critical value $a = a_*$, this branch
of Efimov trimers appears at the atom-dimer threshold:
$E_T^{(n_*)}=E_D$.  As $a$ increases, its binding energy 
$E_T^{(n_*)}-E_D$ relative to the atom-dimer threshold increases 
but its binding energy $E_T^{(n_*)}$ relative to the 3-atom threshold
decreases monotonically. As $a \to \infty$, the binding energy approaches 
a nontrivial limit: $E_T^{(n_*)} \to \hbar^2 \kappa_*^2/m$.
For $a<0$, as $|a|$ decreases, 
$E_T^{(n_*)}$ continues to decrease monotonically.
Finally, at some negative critical value $a=a_*'$,
it disappears through the 3-atom threshold.

%%%%%%%%%%%%%%%%%%%%%%%%%%%%%%%%%%%%%%%%%%%%%%%%%%
%   Binding energies of Efimov states
%%%%%%%%%%%%%%%%%%%%%%%%%%%%%%%%%%%%%%%%%%%%%%%%%%

\subsection{Efimov trimers}
        \label{subsec:B3}

The binding energies of the Efimov states are functions of $a$ and 
$\kappa_*$.  Efimov showed that the calculation of the binding energies 
for all the Efimov states could be reduced to the calculation 
of a single universal function of $\xi$.  The Efimov states can be 
interpreted as bound states in the lowest adiabatic hyperspherical
potential. This potential has a scale-invariant region
where the general solution is the sum of an outgoing hyperradial wave and an
incoming hyperradial wave as in Eq.~(\ref{f-general}).
Bound states occur at energies for which the wave reflected from the 
long-distance region $R \sim |a|$ come into resonance 
with the wave reflected from the
short-distance region $R \sim \ell$.  The resonance condition 
can be expressed as
%----------------------
\begin{eqnarray}
2\theta_* +  \Delta (\xi)=0 \mod 2\pi \,,
\label{theta-Delta}
\end{eqnarray}
%----------------------
%[H,E]
where $\Delta (\xi)/2$ is the phase shift of a hyperradial wave 
that is reflected from the long-distance region.
Using the expression for $\theta_*$ in Eq.~(\ref{theta-star}) and the
definitions for $H$ and $\xi$ in Eq.~(\ref{Hxi-def}), 
we obtain Efimov's equation for the binding energies:
%----------------------
\begin{eqnarray}
E_T  + {\hbar^2 \over m a^2} =
\left(e^{-2 \pi/ s_0} \right)^{n-n_*}
\exp \left[ \Delta ( \xi )/s_0 \right] \frac{\hbar^2\kappa_*^2}{m} \,,
%\nonumber\\
\label{B3-Efimov}
\end{eqnarray}
%----------------------
%[E,H]
where the angle $\xi$ is defined by
%----------------------
\begin{eqnarray}
\label{xin-def}
\tan \xi = - (mE_T/\hbar^2)^{1/2} \, a  \,.
\end{eqnarray}
%----------------------
%[H,E]
We have absorbed the constant $c$ in Eq.~(\ref{theta-star})
into the function $\Delta ( \xi )$ so that it satisfies
$\Delta( -{1\over2}\pi )=0$.

Once the universal function $\Delta(\xi)$ has been calculated, 
the binding energies for all the Efimov states for any values of
$a$ and $\kappa_*$ can be obtained by solving Eq.~(\ref{B3-Efimov}).
The equation is the same for different Efimov states except for the
factor of $(e^{-2 \pi/s_0})^n$ on the right side. 

Efimov's universal function $\Delta(\xi)$ was
calculated in Ref.~\cite{BHK-02a} with a few digits of precision 
over the entire range $-\pi < \xi < - {1 \over 4} \pi$.
It has been calculated in Ref.~\cite{Mohr03} with a precision 
of about 12 digits for $a>0$, which corresponds to the range 
$-{1\over2} \pi< \xi < -{1\over4} \pi$.
It decreases from 6.0273 at $\xi = - {1 \over 4} \pi$
to 0 at $\xi = - {1 \over 2} \pi$ and then to about $-0.89$
at $\xi = - \pi$.
A parameterization of $\Delta(\xi)$ is given in Ref.~\cite{bigrev}
that has errors less than about 0.01, at least in the range
$-{1\over2} \pi< \xi < -{1\over4} \pi$.

In the resonant limit $a \to \pm \infty$,
the spectrum $E_T^{(n)}$ of the Efimov states is particularly simple.
In this limit, $\xi \to - {1\over2} \pi$ and $\Delta(\xi) \to 0$,
so the solutions to Eq.~(\ref{B3-Efimov}) reduce to 
%----------------------
\begin{eqnarray}
E_T^{(n)} = \left( e^{-2 \pi/s_0} \right)^{n-n_*}
{\hbar^2 \kappa_*^2 \over m} \,,
\hspace{1cm} a = \pm \infty,
\label{B3n-n*}
\end{eqnarray}
%----------------------
%[H,E] 
where $\kappa_*$ is the binding wave number for the 
Efimov state labeled by $n = n_*$. 
The spectrum in 
Eq.~(\ref{B3n-n*}) is geometric, with the binding energies of 
successive Efimov states having the ratio 
$e^{-2 \pi / s_0} \approx 1/515.03$. 
The Schr\"odinger wave function in the center-of-mass frame 
for an Efimov state with binding energy 
$E_T=\hbar^2 \kappa^2/m$ is
%----------------------
\begin{eqnarray}
\Psi ({\bm r}_1, {\bm r}_2, {\bm r}_3) = R^{-5/2} f_0 (R) 
\sum_{i = 1}^3 {\sinh \left [ s_0 ({\pi \over 2} - \alpha_i) \right ] 
        \over \sin (2 \alpha_i)} \,.
%\nonumber\\
\end{eqnarray}
%----------------------
%[E]
The hyperradial wavefunction is
%----------------------
\begin{eqnarray}
f_0 (R) = R^{1/2} K_{i s_0} (\sqrt{2} \kappa R) \,,
\label{f0-k}
\end{eqnarray}
%----------------------
%[H,E]
where $K_{i s_0}(z)$ is a Bessel function with imaginary index.
A quantitative measure of the size of a 3-body bound state is 
the mean-square hyperradius: 
%----------------------
\begin{eqnarray}
\langle R^2 \rangle^{(n)} = \left( e^{2 \pi /s_0} \right)^{n-n_*} 
{2(1 + s_0^2)\over 3} 
\kappa_*^{-2} \,.
\end{eqnarray}
%----------------------
%[E]
Thus the root-mean-square hyperradius for each successively shallower Efimov
state is larger than the previous one by $e^{\pi / s_0} \approx 22.7$.

The Efimov trimers disappear through the atom-dimer threshold
at positive critical values of $a$, as illustrated in 
Fig.~\ref{fig:efistar}. For the branch of Efimov trimers 
labelled by $n=n_*$, the critical value is
%----------------------
\begin{eqnarray}
a_*  = 0.0707645086901 \, \kappa_*^{-1} \,.
\label{B3=B2}
\end{eqnarray}
%----------------------
%[E]
The other critical values are $( e^{\pi/s_0} )^n a_*$, 
where $n$ is an integer. 
The binding energy $E_T^{(N)}$ for the Efimov state 
just below the atom-dimer threshold is 
%----------------------
\begin{eqnarray}
E_T^{(N)} \approx E_D \big[ 1 + 0.164\,  \ln^2(a/a_*) \big] \,.
\label{B3approxb}
\end{eqnarray}
%----------------------
%[E,H]
The errors in this approximation scale as $(a-a_*)^3$.

 Efimov states near the 
atom-dimer threshold can be understood intuitively 
as 2-body systems composed of an atom
of mass $m$ and a dimer of mass $2m$.
We can exploit the universality of the 2-body systems 
with large scattering lengths to deduce some properties of the
shallowest Efimov state when it is close to the atom-dimer threshold.
The analog of the universal formula in Eq.~(\ref{B2-uni}) 
is obtained by replacing the reduced mass $m/2$ of the atoms 
by the reduced mass $2m/3$ of the atom and dimer
and by repacing $a$ by the atom-dimer scattering length $a_{AD}$
which diverges at $a=a_*$.
Thus the binding energy relative to the 3-atom threshold 
can be approximated by
%----------------------
\begin{eqnarray}
E_T^{(N)}  \approx E_D + {3 \hbar^2 \over 4m a_{AD}^2} \,.
\label{B3approxa}
\end{eqnarray}
%----------------------
%[H,E]
An explicit expression for the atom-dimer scattering length $a_{AD}$
is given in Eq.~(\ref{a12-explicit}).  The binding energy in 
Eq.~(\ref{B3approxa}) agrees with 
the result in Eq.~(\ref{B3approxb}) up to errors of order 
$a/a_{AD}^3$, which scales as $(a-a_*)^3$.
We can also use universality to deduce the wave function 
of the Efimov trimer.  
The Schr\"odinger wave function  can be expressed as the sum of
three Faddeev wave functions as in Eq.~(\ref{psi3}). 
In the limit $r_{1,23} \gg r_{23}$, 
the first Faddeev wave function should have the form
%----------------------
\begin{eqnarray}
\psi^{(1)}({\bm r}_{23}, {\bm r}_{1,23}) 
\approx \psi_{AD}(r_{1,23}) \, \psi_D (r_{23}) \,, 
\end{eqnarray}
%----------------------
%[E,H]
where $\psi_D(r)$ is the dimer wave function given in Eq.~(\ref{psi-D}) 
and $\psi_{AD}(r)$ is the analogous universal wave function 
for a shallow bound state consisting of two particles 
with large positive scattering length $a_{AD}$:
%----------------------
\begin{eqnarray}
\psi_{AD}(r) = {1 \over r} e^{-r/a_{AD}} \,. 
\label{psi2-AD}
\end{eqnarray}
%----------------------
%[E,H]
This Faddeev wave function can be expressed in terms of 
hyperspherical coordinates using Eqs.~(\ref{hyperradius})
and (\ref{alphak}).
Most of the support of the probability density $|\Psi|^2$ 
is concentrated in the region in which the hyperradius is very large,
$R \sim a_{AD}$, and one of the three   hyperangles is very small,
$\alpha_i \ll 1$.  The mean-square hyperradius can be calculated 
easily when $a_{AD} \gg a$:
%----------------------
\begin{eqnarray}
\langle R^2 \rangle^{(N)} \approx \mbox{$1 \over 3$} a_{AD}^2 \,. 
\end{eqnarray}
%----------------------
%[E]
This result can be obtained more easily simply by using the 
universal atom-dimer wave function in Eq.~(\ref{psi2-AD})
and the approximate expression $R^2 \approx {2\over3} r^2$
for the hyperradius.

The Efimov trimers disappear through the 3-atom threshold
at negative critical values of $a$, as  illustrated in 
Fig.~\ref{fig:efistar}.  For the branch of Efimov trimers 
labelled by $n=n_*$, the critical value is
%----------------------
\begin{eqnarray}
a_*' = - 1.56(5) \,\kappa_*^{-1} \,.
\label{B3=0}
\end{eqnarray}
%----------------------
%[H]
The other critical values are $( e^{\pi/s_0} )^n a_*'$, 
where $n$ is an integer.
In contrast to $a_*$ in Eq.~(\ref{B3=B2}), only a few of digits of precision 
are currently available for $a_*'$.
Comparing Eqs.~(\ref{B3=B2}) and (\ref{B3=0}), we see that 
$a_*'  \approx - 22.0 \;  a_*$.

There is an Efimov trimer at the 3-atom threshold when $a$ has the 
negative critical value $a_*'$.  As can be seen in 
Fig.~\ref{fig:efistar}, the binding energy $E_T^{(n)}$ of the 
Efimov trimer increases rapidly as a function of $a$ 
when $|a|$ exceeds $|a_*'|$.  The increase is so rapid that in 
Ref.~\cite{BHK-02a}, a parameterization of the universal function
$\Delta(\xi)$ in Eq.~(\ref{B3-Efimov}) with an essential singularity at 
$a = a_*'$ was used to get a good fit to the binding energy.
The result of Refs.~\cite{Jonsell06} and \cite{YFT06}
seem to indicate that $E_T^{(n)}$ is actually linear in $1/a$ near 
$a = a_*'$ with a large slope.  The results of a calculation in 
Ref.~\cite{YFT06} can be used to obtain the approximation
%----------------------
\begin{eqnarray}
E_T^{(n)} = \left( 1.1 \, \ln {a \over a_*'} \right)
{\hbar^2 \over  m a^2}  \,.
\label{Efimov0}
\end{eqnarray}
%----------------------

When $|a|$ decreases below $|a_*'|$, the Efimov trimer does not 
immediately disappear, but instead becomes a resonance that decays 
into three atoms.  This resonance is associated with a pole at 
a complex energy $E_{\rm res} - i \Gamma_{\rm res}/2$ with a 
positive real part and a negative imaginary part.  
This complex energy is the analytic continuation of the energy
$-E_T^{(n)}$ of the shallowest Efimov trimer for  $|a| < |a_*'|$.
For $a$ close to $a_*'$, the real part of the resonance energy
is given simply by the negative of the
expression in Eq.~(\ref{Efimov0}).
The authors of Ref.~\cite{YFT06} gave a parameterization 
of the imaginary part of the resonance energy
that scales as $(a-a_*')^2$ near the threshold.  
However their numerical results seem to suggest that 
$\Gamma_{\rm res}$ scales as a higher power of $a-a_*'$.  

There is a common misconception in the literature 
that Efimov states must have binding energies that differ 
by multiplicative factors of 515.03.  However, this
ratio applies only in the resonant limit $ a \to \pm \infty$.  
The ratio $E_T^{(n-1)}/E_T^{(n)}$ of the binding energies
of adjacent Efimov trimers can be much smaller than 515 
if $a > 0$ and much larger than 515 if $a < 0$.
The smallest ratios occur at the critical values $( e^{\pi/s_0} )^n a_*$,
where $a_*$ is given in Eq.~(\ref{B3=B2}). 
The accurate results of Ref.~\cite{Mohr03} for the 
binding energies $E_T$ of the first few Efimov states in units of 
$E_D = \hbar^2/ma^2$ are 
%----------------------
\begin{subequations}
\begin{eqnarray}
E_T^{(N)} \hspace{0.3cm} &=& E_D \,,
\\
E_T^{(N-1)} &=& 6.75029015026 \, E_D \,,
\\
E_T^{(N-2)} &=& 1406.13039320 \, E_D \,.
\end{eqnarray}
\end{subequations}
%----------------------
%[E]
Thus, if $a>0$,  the ratio $E_T^{(N-1)}/E_T^{(N)}$ of the binding energies
for the two shallowest Efimov trimers
can range from about 6.75 to about 208.
The largest ratios occur at the critical values 
$( e^{\pi/s_0} )^n a_*'$, where  $a_*'$ is given in Eq.~(\ref{B3=0}). 
The binding energies $E_T$ of the first few Efimov states are \cite{BHK-02a}
%----------------------
\begin{subequations}
\begin{eqnarray}
E_T^{(N)} \hspace{0.3cm} &=& 0 \,,
\\
E_T^{(N-1)} &=& 1.09 \times 10^3 \; \hbar^2/ma^2 \,,
\\
E_T^{(N-2)} &=& 5.97 \times 10^5 \; \hbar^2/ma^2 \,.
\end{eqnarray}
\end{subequations}
%----------------------
%[H]
Thus, if $a<0$,
the ratio $E_T^{(N-1)}/E_T^{(N)}$ of the binding energies 
for the two shallowest Efimov states can range from about 
550 to $\infty$.

%%%%%%%%%%%%%%%%%%%%%%%%%%%%%%%%%%%%%%%%%%%%%%%%%%
%    Atom-dimer elastic scattering
%%%%%%%%%%%%%%%%%%%%%%%%%%%%%%%%%%%%%%%%%%%%%%%%%%

\subsection{Atom-dimer elastic scattering}
        \label{sec:atom-dimer}

The differential cross section for elastic atom-dimer scattering 
near the atom-dimer threshold can be expressed in terms of the 
atom-dimer scattering length $a_{AD}$:
%----------------------
\begin{eqnarray}
{d \sigma_{AD} \over d \Omega}
\longrightarrow |a_{AD}|^2
\,,\qquad {\rm as\ } E \to - E_D \,.
\label{sigAD}
\end{eqnarray}
%----------------------
The discrete scaling symmetry implies that $a_{AD}/a$ must be 
a log-periodic function of $a \kappa_*$ 
with period $\pi/s_0$.  Its functional form was deduced 
by Efimov up to a few numerical constants. The numerical 
constants were first calculated in Ref.~\cite{Sim81}.
The atom-dimer scattering length is
%----------------------
\begin{eqnarray}
a_{AD} = \big( 1.46+2.15 \cot [s_0 \ln (a/a_*)] \big) \; a \,,
\label{a12-explicit}
\end{eqnarray}
%----------------------
%[e!,h!]
where $a_*\approx 0.071 \, \kappa_*^{-1}$ is given 
to high precision in Eq.~(\ref{B3=B2}).
The atom-dimer scattering length diverges if $a$ has one of the values
$(e^{\pi/s_0})^n a_*$ for which there is an 
Efimov state at the atom-dimer threshold.
It vanishes if $a$ has one of the values
$(e^{\pi/s_0})^n \, 0.38 \, a_*$.

%%%%%%%%%%%%%%%%%%%%%%%%%%%%%%%%%%%%%%%%%%%%%%%%%%
%    Three-body recombination
%%%%%%%%%%%%%%%%%%%%%%%%%%%%%%%%%%%%%%%%%%%%%%%%%%

\subsection{Three-body recombination}
        \label{sec:3BR}

Three-body recombination is a process in which three atoms collide to form
a diatomic molecule and an atom. The energy released 
by the binding energy of the molecule goes into the
kinetic energies of the molecule and the recoiling atom.
The 3-body recombination rate depends on the momenta of
the three incoming atoms. If their momenta are sufficiently small
compared to $1/|a|$, the dependence on the momenta can be neglected,
and the recombination rate reduces to a constant. 
The {\it recombination event rate constant} $\alpha$ is defined 
such that the number of
recombination events per time and per volume in a gas of cold
atoms with number density $n_A$ is $\alpha n^3_A$. 
If the atom and the dimer produced by the recombination process 
have large enough kinetic energies to escape from the system,
the rate of decrease in the number density of atoms is
%----------------------
%\begin{subequations}
\begin{eqnarray}
{d \ \over dt} n_A
& = & - 3 \alpha n^3_A \,,
\label{dnA-gas}
%\\
%& = & - \mbox{$1 \over 2$} \alpha n_A^3 \hspace{1cm} {\rm (BEC)} \,.
%\label{dnA-BEC}
\end{eqnarray}
%\label{dnA-3br}
%\end{subequations}
%----------------------
%[H,E]

In a Bose-Einstein condensate, the three atoms must
all be in the same quantum state, so the coefficient of $n_A^3$ in 
Eq.~(\ref{dnA-gas}) must be multiplied by 1/3! to
account for the symmetrization of the
wave functions of the three identical particles \cite{KSS-85}.
This prediction was first tested by the JILA group~\cite{Burt97}.
They measured the 3-body loss rates 
in an ultracold gas of $^{87}$Rb atoms in the 
$|f,m_f\rangle=|1,-1\rangle$ hyperfine state,
both above and below the critical temperature for  
Bose-Einstein condensation. They 
found that the loss rate was smaller in the Bose-Einstein condensate
by a factor of $7.4(2.6)$, in agreement with the  
predicted value of 6 \cite{KSS-85}. 

If the scattering length $a$ is negative, 
the molecule can only be a deep diatomic molecule 
with binding energy of order $\hbar^2/m \ell^2$ or
larger. However, if $a$ is positive and unnaturally large ($a \gg \ell$),
the molecule can also be the shallow dimer with binding
energy $E_D = \hbar^2/ma^2$. Three-body recombination into deep 
dimers will be discussed in Section~\ref{sec:deep}.
In this section, we assume there are no deep dimers.
We therefore assume $a>0$ and focus on 3-body
recombination into the shallow dimer. 
We denote the contribution to the rate constant $\alpha$ 
from 3-body recombination into the shallow dimer by $\alpha_{\rm shallow}$.

Dimensional analysis together with the discrete scaling symmetry 
implies that $\alpha_{\rm shallow}$ is proportional to $\hbar a^4/m$ 
with a coefficient that is a log-periodic function of $a\kappa_*$
with period $\pi/s_0$.
An analytic expression for  $\alpha_{\rm shallow}$ has 
recently been derived \cite{Petrov-octs,MOG05}:
%-----------------
\begin{eqnarray}
\alpha_{\rm shallow} = 
{128 \pi^2 (4 \pi - 3 \sqrt{3}) \sin^2 [s_0 \ln (a/a_{*0})]\over
\sinh^2(\pi s_0) + \cos^2 [s_0 \ln (a/a_{*0})]} \,
{\hbar a^4 \over m} \,,
\label{alpha-analytic}
\end{eqnarray}
%----------------------
%[e!,h!]
where $a_{*0}$ differs from $\kappa_*^{-1}$ by a multiplicative constant 
that is known only to a couple of digits of accuracy:
%----------------------
\begin{eqnarray}
a_{*0}  \approx 4.5 \, a_*  \approx 0.32 \, \kappa_*^{-1} \,.
\label{a*0}
\end{eqnarray}
%----------------------
%[e!,h!]
The maximum value of the coefficient of $\hbar a^4/m$ 
in Eq.~(\ref{alpha-analytic}) is 
%----------------------
\begin{eqnarray}
C_{\rm max} = 
{128 \pi^2 (4 \pi - 3 \sqrt{3}) \over\sinh^2(\pi s_0)} \,.
\label{C-max}
\end{eqnarray}
%----------------------
%[E]
Its numerical value is $C_{\rm max} = 67.1177$.
We can exploit the fact that 
$\sinh^2(\pi s_0) \approx 139$ is large
to simplify the expression in Eq.~(\ref{alpha-analytic}).
The rate constant can be approximated 
with an error of less than 1\% of $C_{\rm max}\hbar a^4/m$ by
%----------------------
\begin{eqnarray}
\alpha_{\rm shallow} \approx 
67.12 \sin^2 [s_0 \ln (a/a_{*0}) ] \, \hbar a^4/m \,.
\label{alpha-sh}
\end{eqnarray}
%----------------------
%[e!,h!]
This approximate functional form of the rate constant was first deduced
in Refs.~\cite{NM-99,EGB-99}.
The coefficient $C_{\rm max}$ and the relation between $a_{*0}$ 
and $\kappa_*$ was calculated accurately in Refs.~\cite{BBH-00,BH02}.

The most remarkable feature of the analytic expression in 
Eq.~(\ref{alpha-analytic}) and the approximate expression 
in Eq.~(\ref{alpha-sh}) is that the coefficient of $\hbar a^4/m$ 
oscillates between zero and 67.12 as a function of $\ln(a)$.
In particular, $\alpha_{\rm shallow}$ has zeroes at values of $a$ 
given by $( e^{\pi/s_0})^n \, a_{*0}$, where 
$a_{*0} \approx 4.5 \, a_*$. The maxima of $\alpha_{\rm shallow}$  
in Eq.~(\ref{alpha-sh}) occur at the values of $a$
for which $\tan[s_0 \ln (a/a_{*0})]= - s_0/2$, which are 
$a\approx ( e^{\pi/s_0} )^n \, 64.3 \, a_* \approx 
( e^{\pi/s_0} )^n 4.6  \,\kappa_*^{-1}$.

Since the zeroes in $\alpha_{\rm shallow}$ are so remarkable, it is worth
enumerating the effects that will tend to fill in the zeroes, 
turning them into local minima of $\alpha_{\rm shallow}$. 
The zeroes arise from the interference between two pathways
associated with the lowest two adiabatic hyperspherical channels.
There may be additional contributions from coupling to higher 
hyperspherical channels, but in the scaling limit
they are strongly suppressed numerically.
The zero in $\alpha_{\rm shallow}$ for $a=a_{*0}$ is exact 
only at threshold. If the recombining atoms have 
collision energy $E$, $\alpha_{\rm shallow}$ 
goes to zero as $E^2$ as $E \rightarrow 0$ \cite{SEGB02}.
Thus thermal effects that give nonzero collision energy to the atoms 
will tend to fill in the zeroes.  
Finally, if the 2-body potential supports deep
diatomic molecules, their effects will tend to fill in the zeros of 
$\alpha_{\rm shallow}$ as described in Section~\ref{sec:deep-3BR}.  
Furthermore, as described in Section~\ref{sec:deep-3BR},
3-body recombination into those deep dimers gives an additional
nonzero contribution $\alpha_{\rm deep}$ to the rate constant $\alpha$.

%\newpage

%%%%%%%%%%%%%%%%%%%%%%%%%%%%%%%%%%%%%%%%%%%%%%%%%%
%
%     EFFECTS OF DEEP TWO-BODY BOUND STATES
%
%%%%%%%%%%%%%%%%%%%%%%%%%%%%%%%%%%%%%%%%%%%%%%%%%%

\section{Effects of Deep Diatomic Molecules}
        \label{sec:deep}

In this Section, we describe the effects of deep diatomic 
molecules on the universal properties for three identical bosons 
in the scaling limit.

%%%%%%%%%%%%%%%%%%%%%%%%%%%%%%%%%%%%%%%%%%%%%%%%%%
%    Boundary condition at short distances
%%%%%%%%%%%%%%%%%%%%%%%%%%%%%%%%%%%%%%%%%%%%%%%%%%

\subsection {Boundary condition at short distances}
        \label{sec:deep-RL}

The existence of deep dimers requires a modification of the boundary 
condition on the hyperradial wave function at short distances.
The general solution for the hyperradial wave function 
in the lowest hyperspherical potential in the
scale-invariant region $\ell \ll R \ll |a|$ is given in Eq.~(\ref{f-general}).
The boundary condition in Eq.~(\ref{AB}), which takes into account 
the effects of short distances $R \lsim \ell$, 
follows from the assumption that all the probability
in an incoming hyperradial wave is reflected back from the short-distance
region into an outgoing hyperradial wave. 
If there are deep dimers, 
some of the probability in the incoming hyperradial
wave that flows into the short-distance region emerges in the form of
scattering states that consist of an atom and a deep dimer
with large kinetic energy but small total energy.
We will refer to these states as 
{\it high-energy atom-dimer scattering states}.

The 2-body potentials for the alkali atoms other than hydrogen 
support diatomic molecules with many vibrational energy levels.
If it was necessary to take into account each of
these deep dimers explicitly, the problem would be extremely difficult.
Fortunately the cumulative effect of all the deep dimers on
low-energy 3-atom observables can be taken into account through
one additional parameter: an inelasticity parameter $\eta_*$
that determines the widths of the Efimov trimers. 
In the scaling limit, the low-energy 3-body
observables are completely determined by $a$, $\kappa_*$, and $\eta_*$.

The reason the cumulative effects of the deep dimers can be
described by a single number $\eta_*$ is that all pathways from 
a low-energy 3-atom state to a high-energy
atom-dimer scattering state must flow through the lowest 
hyperspherical potential, which in the scale-invariant region has the
form given in Eq.~(\ref{V0-si}).
The reason for this is that in order to reach a high-energy atom-dimer
scattering state, the system must pass through an intermediate 
configuration in which all three atoms are simultaneously close together
with a hyperradius $R$ of order $\ell$ or smaller. Such small values of
$R$ are accessible to a low-energy 3-atom state
only through the lowest hyperspherical potential.

If there are deep dimers, some of the probability in a hyperradial wave
that flows to short distances emerges in the form 
of high-energy atom-dimer scattering states.  We denote 
the fraction of the probability that is reflected back to long distances
by $e^{-4 \eta_*}$.
We refer to $\eta_*$ as the Efimov width parameter.  
The amplitude $A$ of the outgoing wave in Eq.~(\ref{f-general})
then differs from the amplitude $B$ of the incoming wave
not only by a phase as in Eq.~(\ref{AB}) but also by a suppression factor
$e^{-2\eta_*}$.  Thus if there are deep 2-body bound states, 
the boundary condition in Eq.~(\ref{AB}) must be replaced by
%----------------------
\begin{eqnarray}
A = - e^{-2 \eta_* + 2 i \theta_*} B.
\label{AB-deep}
\end{eqnarray}
%----------------------
%[E]
The angle $\theta_*$ is related to the 3-body parameter $\kappa_*$ 
by Eq.~(\ref{theta-star}).  The parameters $\kappa_*$ and $\eta_*$
appear in the boundary condition in Eq.~(\ref{AB-deep}) 
in the combination $\theta_* + i \eta_*$.  If the universal results
for the case $\eta_*=0$ can be expressed as an analytic function 
of $\ln(\kappa_*)$, then the corresponding universal results
for the case in which there are deep 2-body bound states
can be obtained simply by the substitution 
$\ln(\kappa_*) \to \ln(\kappa_*) + i \eta_*/s_0$.

One can also obtain universal results 
for the total probability of transitions from low-energy 3-atom 
or atom-dimer scattering states to 
high-energy atom-dimer scattering states.
The transition rate for any individual 
deep dimer is sensitive to the details of
wave functions in the short-distance region $R \sim \ell$.  
However the inclusive rate summed over all 
deep dimers is much less sensitive to 
short distances, because it is constrained by 
probability conservation.  The inclusive rates are given by 
universal expressions that include a factor of
$1 - e^{-4 \eta_*}$ \cite{Braaten:2003yc},
which is the probability that a hyperradial wave is not 
reflected from the short-distance region.

%%%%%%%%%%%%%%%%%%%%%%%%%%%%%%%%%%%%%%%%%%%%%%%%%%
%    Widths of Efimov states
%%%%%%%%%%%%%%%%%%%%%%%%%%%%%%%%%%%%%%%%%%%%%%%%%%

\subsection {Widths of Efimov states}
        \label{sec:deep-widths}

One obvious consequence of the existence of deep dimers 
is that the Efimov trimers are no longer sharp states. 
They are resonances with nonzero widths, 
because they can decay into an atom and a deep dimer.    
The binding energy $E_T$ and width $\Gamma_T$ of an Efimov resonance 
can be obtained as a complex eigenvalue $E=-(E_T+i\Gamma_T/2)$
of the 3-body Schr{\"o}dinger equation.  

In the absence of deep dimers, 
the binding energies of Efimov trimers satisfy Eq.~(\ref{B3-Efimov}), 
where $\Delta(\xi)/2$ is the phase shift of a hyperradial wave that is
reflected from the long-distance region $R\sim |a|$.  To obtain the
corresponding equation in the case of deep dimers, we need only
make the substitution $ \theta_* \to \theta_* + i \eta_*$ in
Eq.~(\ref{theta-Delta}):
%----------------------
\begin{eqnarray}
2(\theta_* + i\eta_*) + \Delta(\xi) = 0 \mod 2 \pi.
\end{eqnarray}
%----------------------
%[E,H]
This can be satisfied only if we allow complex values of $\xi$
in the argument of $\Delta$. Using the expression
for $\theta_*$  in Eq.~(\ref{theta-star}) and inserting the definition 
of $H$ in Eq.~(\ref{Hxi-def}), we obtain the equation
%----------------------
\begin{eqnarray}
E_T+{i\over 2}\Gamma_T + {\hbar^2 \over ma^2} 
%&&  \nonumber\\ && \hspace{-3cm}
&=& 
\left( e^{-2\pi /s_0} \right)^{n-n_*} 
\exp \left[ {\Delta(\xi)+ 2 i \eta_* \over s_0} \right] 
{\hbar^2 \kappa_*^2 \over m} \,,
\label{B3-Efimov-width}
\end{eqnarray}
%[E,H]
%----------------------
where the complex-valued angle $\xi$ is defined by
%----------------------
\begin{eqnarray}
\tan \xi = - \big( m(E_T + i \Gamma_T/2)/\hbar^2 \big)^{1/2} \, a \,.
\end{eqnarray}
%----------------------
%[E,H]
To solve this equation for $E_T$ and $\Gamma_T$, we need the analytic
continuation of
$\Delta(\xi)$ to complex values of $\xi$.
The parametrization for $\Delta(\xi)$ in Ref.~\cite{bigrev}
should be accurate for complex values of $\xi$ with sufficiently small
imaginary parts, except near $\xi=-\pi$ where
the parametrization used an expansion parameter 
that was not based on any analytic understanding of the  behavior
near the endpoint.
If the analytic continuation of $\Delta(\xi)$ were known,
the binding energy and width of one Efimov state
could be used to determine $\kappa_*$ and $\eta_*$.
The remaining Efimov states and their widths could then be
calculated by solving Eq.~(\ref{B3-Efimov-width}).

If the Efimov width parameter $\eta_*$ is extremely small, the right
side of Eq.~(\ref{B3-Efimov-width}) can be expanded to first order in $\eta_*$.
The resulting expression for the width is 
%----------------------
\begin{eqnarray}
\Gamma_T \approx {4 \eta_*\over s_0} 
    \left( E_T +{\hbar^2\over ma^2} \right) \,.
\end{eqnarray}
%----------------------
%[E,H]
For the shallowest Efimov states, the order of magnitude of the width is simply
$\eta_* \hbar^2 /ma^2$.  The widths of the deeper Efimov states are
proportional to their binding energies, which behave asymptotically like
Eq.~(\ref{B3-resonant}).  This geometric increase in the widths of 
deeper Efimov states has been observed in calculations 
of the elastic scattering of atoms with deep dimers \cite{NSE00}.

%%%%%%%%%%%%%%%%%%%%%%%%%%%%%%%%%%%%%%%%%%%%%%%%%%
%   Atom-dimer scattering
%%%%%%%%%%%%%%%%%%%%%%%%%%%%%%%%%%%%%%%%%%%%%%%%%%

\subsection{Atom-dimer scattering}
        \label{sec:deepAD}

The effects of deep dimers modify the universal 
expressions for low-energy 3-body scattering observables
derived in Section~\ref{sec:uni3}. If the universal 
expression for a scattering amplitude for the case of no deep dimers 
is expressed as an analytic function of $\ln(\kappa_*)$,
the corresponding universal expression for the case in which there 
are deep dimers can be obtained simply by substituting
$\ln(\kappa_*) \to \ln(\kappa_*) + i \eta_*/s_0$.

\begin{figure}[htb]
\centerline{\includegraphics*[width=8.5cm,angle=0,clip=true]{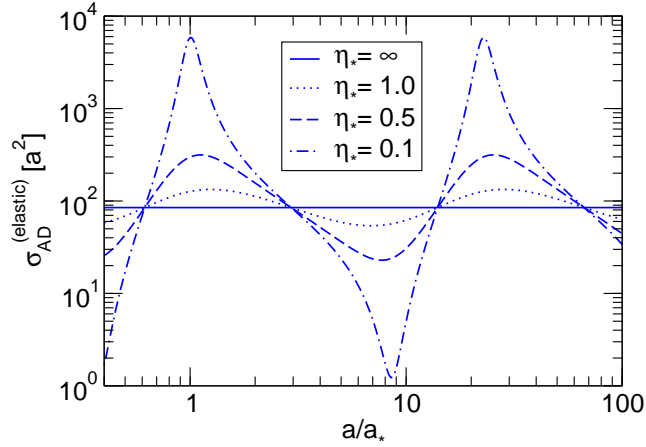}}
\vspace*{0.0cm}
\caption{The elastic cross section for atom-dimer scattering at threshold
in units of $a^2$ as a function of $a/a_*$
for several values of $\eta_*$.}
\label{fig:ddadel}
\end{figure}

The differential cross section for elastic atom-dimer scattering
near the atom-dimer threshold is still given by Eq.~(\ref{sigAD}), 
except that the atom-dimer scattering length
is now complex valued:
%----------------------
\begin{eqnarray}
a_{AD} &=& \big( 1.46
 +2.15 \cot [s_0 \ln (a/a_*) + i \eta_*]
\big) \; a \,, 
\label{aAD-deep}
\end{eqnarray}
%----------------------
%[H,E]
where $a_*\approx 0.071 \kappa_*^{-1}$ is given in Eq.~(\ref{B3=B2}).
The elastic cross section reduces to
\begin{eqnarray}
\sigma_{AD}^{\rm (elastic)}(E = - E_D) = 84.9\,
{\sin^2 [s_0 \ln (a/a_*)+ 0.97] + \sinh^2 \eta_* 
\over \sin^2 [s_0 \ln (a/a_*)] + \sinh^2 \eta_*} 
\; a^2  \,.
\label{sigmaAD-deep}
\end{eqnarray}
%----------------------
%[e!,h!]
The coefficient of $a^2$ is shown in Fig.~\ref{fig:ddadel}
as a function of $a$ for several values of $\eta_*$.
In the limit $\eta_* \to \infty$, the log-periodic dependence on 
$a \kappa_*$ disappears and the cross section reduces to $84.9 \, a^2$.

%%%%%%%%%%%%%%%%%%%%%%%%%%%%%%%%%%%%%%%%%%%%%%%%%%
%%%%%%%%%%%%%%%%%%%%%%%%%%%%%%%%%%%%%%%%%%%%%%%%%%

If there are no deep dimers, 
atom-dimer scattering is completely elastic below the 
dimer-breakup threshold $ka = 2/\sqrt{3}$.
The existence of deep dimers opens up inelastic channels 
in which an atom and a shallow dimer with low energy
collide to form an atom and a deep dimer.
The large  binding energy of the deep dimer is released through the  
large kinetic energies of the recoiling atom and dimer. 
This process is called {\it dimer relaxation}.
The optical theorem implies that the total cross section
for inelastic atom-dimer scattering near the atom-dimer threshold 
$E = - E_D$ is proportional to the imaginary part of the atom-dimer 
scattering length in Eq.~(\ref{aAD-deep}):
%----------------------
\begin{eqnarray}
\sigma_{AD}^{\rm (inelastic)}(E\to - E_D)  &\longrightarrow&
\frac{4\pi}{k}\;(-{\rm Im} \, a_{AD}) \,.
\label{ksig-inelastic}
\end{eqnarray}
%----------------------
%[e!,h!]
Thus the cross section for dimer relaxation diverges
like $1/k$ as $E$ approaches the atom-dimer threshold. 

The event rate $\beta$ for {\it dimer relaxation} 
is defined so that the
number of dimer relaxation events per time and per volume 
in a gas of atoms with number density $n_A$ 
and dimers with number density $n_D$ is $\beta n_A n_D$.  
If the atom and the deep dimer produced by the relaxation 
process have large enough kinetic energies to escape 
from the system, the rate of decrease in the number densities is 
%----------------------
\begin{eqnarray}
{d \ \over d t} n_A & = &  {d \ \over d t} n_D
= - \beta n_A n_D \,.
\label{dn-deact}
\end{eqnarray}
%----------------------
The event rate $\beta$ can be expressed in terms of a
statistical average of the inelastic atom-dimer cross section:
%----------------------
\begin{eqnarray}
\beta = {3 \hbar \over 2 m}
\left\langle k \ \sigma_{AD}^{\rm (inelastic)}(E) \right\rangle \,.
\label{beta-T}
\end{eqnarray}
%----------------------
In the low-temperature limit, $\beta$
reduces to $6 \pi \hbar(-{\rm Im} \, a_{AD})/m$,
which can be written as \cite{Braaten:2003yc}
%----------------------
\begin{eqnarray}
\beta &=&
{20.3 \sinh(2\eta_*) \over \sin^2 [s_0 \ln (a/a_*)] + \sinh^2 \eta_*}
\; {\hbar a \over m}  \,.
\label{beta:a>0}
\end{eqnarray}
%----------------------
If $\eta_*$ is small, the maximum value of $\beta$ occurs when
$a \approx [1 + \sinh^2 \eta_*/(2 s_0^2)] a_*$.

%%%%%%%%%%%%%%%%%%%%%%%%%%%%%%%%%%%%%%%%%%%%%%%%%%
\begin{figure}[htb]
\centerline{\includegraphics*[width=8.5cm,angle=0,clip=true]{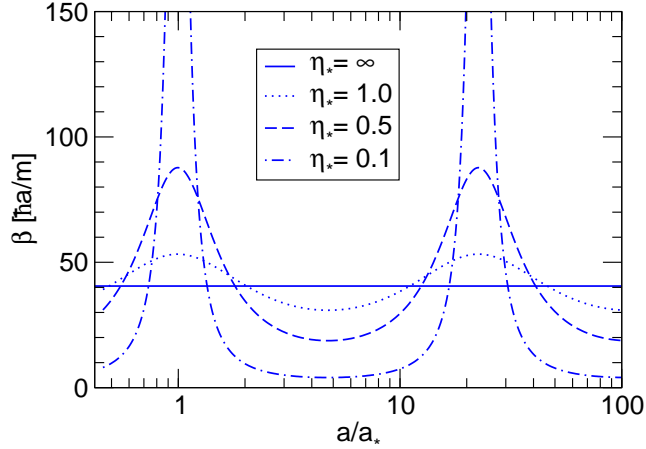}}
\vspace*{0.0cm}
\caption{The dimer relaxation rate constant
in units of $\hbar a/m$ as a function of $a/a_*$
for several values of $\eta_*$.
}
\label{fig:ddad}
\end{figure}
%%%%%%%%%%%%%%%%%%%%%%%%%%%%%%%%%%%%%%%%%%%%%%%%%%

The coefficient of $\hbar a/m$ is shown in Fig.~\ref{fig:ddad} 
as a function of $a/a_*$ for several values of $\eta_*$.
It displays resonant behavior with maxima when the scattering length
has one of the values $(e^{\pi/s_0})^n a_*$ for which the peak of an 
Efimov resonance is at the atom-dimer threshold. 
In the limit $\eta_* \to 0$, the maximum value $40.6 \coth \eta_*$ 
of the coefficient of $\hbar a/m$ diverges.  
In the  limit $\eta_* \to \infty$, 
the log-periodic dependence of the coefficient on $a \kappa_*$ 
disappears and it reduces to the constant 40.6.

%%%%%%%%%%%%%%%%%%%%%%%%%%%%%%%%%%%%%%%%%%%%%%%%%%
%    Three-body recombination 
%%%%%%%%%%%%%%%%%%%%%%%%%%%%%%%%%%%%%%%%%%%%%%%%%%

\subsection{Three-body recombination}
        \label{sec:deep-3BR}

%%%%%%%%%%%%%%%%%%%%%%%%%%%%%%%%%%%%%%%%%%%%%%%%%%
\begin{figure}[htb]
\bigskip
\centerline{\includegraphics*[width=8cm,angle=0]{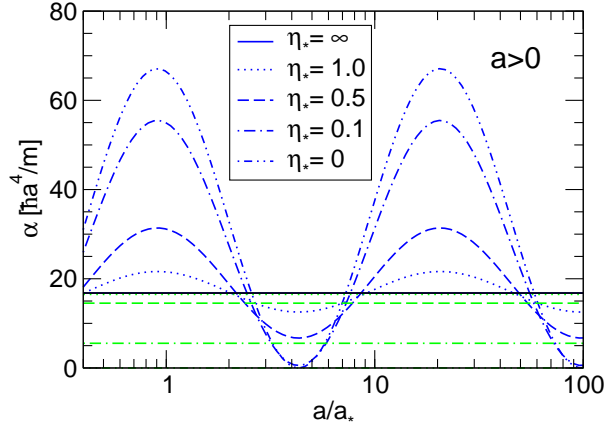}}
\medskip
\caption
{The 3-body recombination rate constants
$\alpha_{\rm shallow}$ (with large-amplitude oscillations) 
and  $\alpha_{\rm deep}$ (with small-amplitude oscillations)
in units of $\hbar a^4/m$ as functions of $a/a_*$ 
for $a >0$ and several values of $\eta_*$.}
\label{fig:alpha-deep+}
\end{figure}
%%%%%%%%%%%%%%%%%%%%%%%%%%%%%%%%%%%%%%%%%%%%%%%%%%

If there are no deep dimers, the rate constant $\alpha_{\rm shallow}$
for 3-body recombination into the shallow dimer 
has the remarkable form given in Eq.~(\ref{alpha-analytic}), 
which has zeroes at values $a$ that differ by multiples of 
$e^{\pi/s_0} \approx 22.7$.  
If there are deep dimers, the analytic expression
for the rate constant $\alpha_{\rm shallow}$  is
%----------------------
\begin{eqnarray}
\alpha_{\rm shallow} = 
{128 \pi^2 (4 \pi - 3 \sqrt{3})
(\sin^2 [s_0 \ln (a/a_{*0})] + \sinh^2\eta_*)
\over
\sinh^2(\pi s_0 + \eta_*) + \cos^2 [s_0 \ln (a/a_{*0})] }\,
{\hbar a^4 \over m} \,,
\label{alpha-analytic:eta}
\end{eqnarray}
%----------------------
%[e!,h!]
where $a_{*0} \approx 4.5 \, a_*$ is given in Eq.~(\ref{a*0}).
We can use the approximation
$\sinh(\pi s_0 + \eta_*) \approx e^{\eta_*} \sinh(\pi s_0)$ 
to simplify the expression in Eq.~(\ref{alpha-analytic:eta}):
%----------------------
\begin{eqnarray}
\alpha_{\rm shallow} &\approx& 67.12 \, e^{-2 \eta_*} 
\left( \sin^2[s_0 \ln (a/a_{*0})] + \sinh^2 \eta_* \right)
\hbar a^4/m .
\label{alpha_sh:deep}
\end{eqnarray}
%----------------------
%[e!,h!]
The coefficient of $\hbar a^4/m$ is shown as a function of $a/a_*$ 
in Fig.~\ref{fig:alpha-deep+} for several values of $\eta_*$.
As $a$ varies, the coefficient of $\hbar a^4 /m$ oscillates between
about $67.12 \, e^{-2 \eta_*} \sinh^2 \eta_*$ 
and about $67.12 \, e^{-2 \eta_*} \cosh^2 \eta_*$. 
Thus one effect of the deep dimers is to eliminate the zeros 
in $\alpha_{\rm shallow}$ at $a=(e^{\pi/s_0})^n a_{*0}$.
The depth of the minimum is quadratic in $\eta_*$
as $\eta_* \to 0$, so the coefficient of $\hbar a^4/m$ 
at $a=(e^{\pi/s_0})^n a_{*0}$ can be very small 
if the Efimov width parameter $\eta_*$ is small.

The existence of deep dimers 
opens up additional channels for 3-body recombination.  
If there are no deep dimers,  3-body recombination 
can only produce the shallow dimer if $a>0$ 
and it cannot proceed at all if $a<0$.
If there are deep dimers, they can be produced by 3-body 
recombination for either sign of $a$.
We denote  by $\alpha_{\rm deep}$ the inclusive contribution to the 
event rate constant defined in Eq.~(\ref{dnA-gas}) from 
3-body recombination into all the deep dimers.

If $a>0$, the analytic expression for $\alpha_{\rm deep}$ is
%----------------------
\begin{eqnarray}
\alpha_{\rm deep} = 
{64 \pi^2 (4 \pi - 3 \sqrt{3}) \coth(\pi s_0) \sinh(2\eta_*)
\over \sinh^2(\pi s_0 + \eta_*) + \cos^2 [s_0 \ln (a/a_{*0})] }\,
{\hbar a^4 \over m} \,,
\label{alpha-deep:eta}
\end{eqnarray}
%----------------------
%[e!,h!]
where $a_{*0} \approx 4.5 \, a_*$ is given in Eq.~(\ref{a*0}).
The coefficient of $\hbar a^4/m$ has very weak 
log-periodic dependence on $a \kappa_*$.
We can use the appproximation 
$\sinh(\pi s_0 + \eta_*) \approx e^{\eta_*} \sinh(\pi s_0)$ 
to simplify the expression in Eq.~(\ref{alpha-deep:eta}):
%----------------------
\begin{eqnarray}
\alpha_{\rm deep} \approx 
16.84 \left( 1 - e^{-4\eta_*} \right) \hbar a^4/m 
\,, \qquad (a>0) \,.
\label{alpha-deep:a>0}
\end{eqnarray}
%----------------------
%[e!,h!]
The numerical result for the coefficient in Eq.~(\ref{alpha-deep:a>0})
was first derived in Ref.~\cite{Braaten:2003yc}.
The coefficient of $\hbar a^4/m$, which is independent of $a/a_*$, 
is shown in Fig.~\ref{fig:alpha-deep+} for several values of $\eta_*$.
In the limit $\eta_* \to 0$, $\alpha_{\rm deep}$ approaches zero
linearly in $\eta_*$.  In the limit $\eta_* \to \infty$, 
the rates $\alpha_{\rm deep}$ in Eq.~(\ref{alpha-deep:eta}) and  
$\alpha_{\rm shallow}$ in Eq.~(\ref{alpha-analytic:eta}) are almost equal,
differing only by the factor $\coth(\pi s_0) \approx 1.004$.

%%%%%%%%%%%%%%%%%%%%%%%%%%%%%%%%%%%%%%%%%%%%%%%%%%
\begin{figure}[htb]
\bigskip
\centerline{\includegraphics*[width=8.5cm,angle=0]{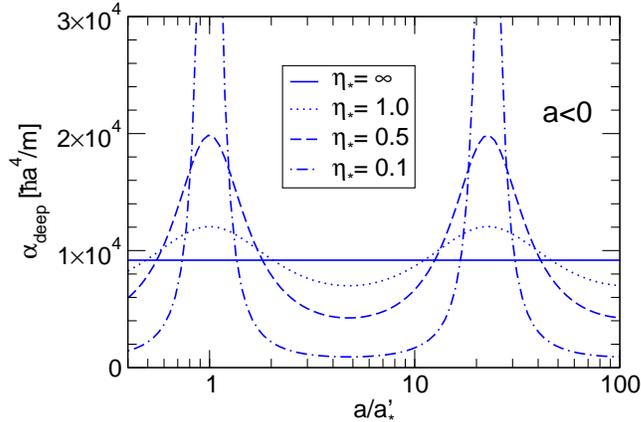}}
\medskip
\caption
{The 3-body recombination rate constant $\alpha_{\rm deep}$ 
in units of $\hbar a^4/m$ as a function of $a/a_*'$ 
for $a<0$ and several values of $\eta_*$.}
\label{fig:alpha-deep-}
\end{figure}
%%%%%%%%%%%%%%%%%%%%%%%%%%%%%%%%%%%%%%%%%%%%%%%%%%

If $a<0$, the 3-body recombination rate constant is
%----------------------
\begin{eqnarray}
\alpha_{\rm deep} &=& 
{4590 \sinh(2 \eta_*) \over \sin^2 [s_0 \ln (a/a_*')] + 
\sinh^2 \eta_*}
\; {\hbar a^4 \over m}\,, \qquad (a<0) \,,
\label{alpha-deep:a<0}
\end{eqnarray}
%----------------------
%[E,H]
where $a_*' \approx -1.6 \, \kappa_*^{-1}$ is given in Eq.~(\ref{B3=0}).
The explicit formula for $\alpha_{\rm deep}$ in Eq.~(\ref{alpha-deep:a<0})
was first derived in Ref.~\cite{Braaten:2003yc}.\footnote{
See also Ref.~\cite{BH01} for an earlier calculation that was
perturbative in $\eta_*$.}
The coefficient 4590 is only known numerically.
The coefficient of $\hbar a^4/m$ is shown as a function of $a/a_*'$ 
in Fig.~\ref{fig:alpha-deep-} for several values of $\eta_*$.
It displays resonant behavior with maxima when the scattering length
has one of the values $(e^{\pi/s_0})^n a_*'$ for which there is an 
Efimov trimer near the 3-atom  threshold. 
In the limit $\eta_* \to 0$, the maximum value $9180 \coth \eta_*$ 
of the coefficient of $\hbar a^4/m$ in Eq.~(\ref{alpha-deep:a<0}) diverges.  
In the  limit $\eta_* \to \infty$, the log-periodic dependence
of the coefficient on $a \kappa_*$ disappears 
and it approaches the constant 9180.

%\newpage

%%%%%%%%%%%%%%%%%%%%%%%%%%%%%%%%%%%%%%%%%%%%%%%%%%
%
%     APPLICATIONS TO ATOMS
%
%%%%%%%%%%%%%%%%%%%%%%%%%%%%%%%%%%%%%%%%%%%%%%%%%%

\section{Applications to Atoms}
        \label{sec:appatom}

In this section, we describe applications of universality
to $^4$He atoms and to alkali atoms near a Feshbach resonance.

%%%%%%%%%%%%%%%%%%%%%%%%%%%%%%%%%%%%%%%%%%%%%%%%%%
%    Helium Atoms
%%%%%%%%%%%%%%%%%%%%%%%%%%%%%%%%%%%%%%%%%%%%%%%%%%

\subsection {Helium atoms}
        \label{sec:he4atoms}

Helium atoms provide a beautiful illustration of universality 
in the 3-body system \cite{BH02}.  The interatomic potential 
between two $^4$He atoms does not support any deep dimers,
so universality is realized in its simplest form with a 
single 3-body parameter $\kappa_*$.
The binding energies of the $^4$He
trimers have been calculated accurately for a number of different model
potentials for the interaction between two $^4$He atoms.  
For the purposes of illustration, we will use the TTY potential \cite{TTY95}.  
The scattering length for the TTY potential is $a = 188.99 \, a_0$.  
This is much larger than its effective range $r_s = 13.85 \, a_0$, 
which is comparable to the van der Waals
length scale $\ell_{\rm vdw} = 10.2 \, a_0$.  
The TTY potential supports a single 2-body bound state, 
the $^4$He dimer whose binding energy is $E_2= 1.30962$ mK.  
This binding energy is small compared to the natural low-energy 
scale for $^4$He atoms:
$E_{\rm vdW} \approx 420$~mK.  
The TTY potential supports exactly two 3-body bound states:  
the ground-state trimer, which we label $n = 0$, 
and the excited trimer, which we label $n = 1$. 
There have been several accurate calculations of the binding energies
$E_3^{(0)}$ and $E_3^{(1)}$ for the TTY potential 
\cite{NFJ98,RY00,MSSK01}.  The results agree to within 
0.5\% for both $E_3^{(0)}$ and $E_3^{(1)}$.  
The results of Ref.~\cite{RY00} are $E_3^{(0)} = 126.4$ mK
and $E_3^{(1)} = 2.28$ mK.

The $^4$He dimer was first observed in 1992 
by the Minnesota group \cite{LMKGG}.
They used the expansion of $^4$He gas at room temperature from a 
pulsed valve into a vacuum chamber to create a beam of $^4$He atoms
and clusters with a translational temperature near 1 mK.
The dimers were detected by using electron impact ionization 
to produce helium dimer ions that were observed by mass spectrometry.
In another experiment in 1995, they determined the size 
of the $^4$He dimer by measuring the relative transmission rates 
of He atoms and dimers through a set of nanoscale sieves \cite{LGG96}.
Their result for 
the mean separation of the atoms in the dimer was 
$\langle r \rangle = (62 \pm 10)$ \AA.

The $^4$He dimer was also detected nondestructively 
by the G\"ottingen group \cite{STo94}.
They made a beam of $^4$He atoms and clusters with temperature 
ranging from 6 K to 60 K by allowing cryogenic $^4$He gas 
to escape from an orifice.  The beam was passed through a 
nanoscale transmission grating, and the dimers were detected 
by observing a diffraction peak at the expected angle.
In 1995, this experiment was used to make the first observation of
the ground state $^4$He trimer \cite{STo96}.
The excited $^4$He trimer has not yet been observed.
In a subsequent experiment, the $^4$He tetramer was observed
and the formation rates of dimers, trimers, and tretramers 
was measured as functions of the temperature and pressure 
of the gas from which the beam escaped \cite{BST02}.
In 2000, the G\"ottingen group determined the size of the dimer 
from the relative strengths of the higher order diffraction peaks 
from the nanoscale transmission grating \cite{Grisenti}. 
Their result for the mean separation of the atoms is 
$\langle r \rangle = (52 \pm 4)$ \AA.
This result is in good agreement with the theoretical 
prediction from the TTY potential:
$\langle r \rangle = a/2 = 50$ \AA.
From this measurement, the binding energy of the $^4$He dimer
was inferred to be $E_2 = (1.1^{+0.3}_{-0.2})$ mK.
The scattering length for $^4$He atoms was inferred 
to be $a = (104^{+8}_{-18})$ \AA.

Lim, Duffy, and Damert proposed in 1977 that the excited state 
of the $^4$He trimer is an Efimov state \cite{LDD77}.
This interpretation is almost universally accepted.
Some researchers have proposed that the ground state trimer 
is also an Efimov state \cite{FTDA99,BHK99,BHK99b}.
This raises an obvious question: 
what is the definition of an Efimov state? 
The most commonly used definition 
is based on rescaling the depth of the 2-body potential:
$V({\bf r})  \longrightarrow \lambda V({\bf r})$.
According to the traditional definition, 
a trimer is an Efimov state if its binding 
energy as a function of the scaling parameter $\lambda$
has the qualitative behavior illustrated in Fig.~\ref{fig:efistar}.
As $\lambda$ is decreased below 1, the trimer eventually 
disappears through the 3-atom threshold.
As $\lambda$ is increased above 1, the trimer eventually 
disappears through the atom-dimer threshold.  
Calculations of the trimer binding energies \cite{Esry96a}
using a modern helium potential show that the excited trimer 
satisfies this definition of an Efimov state
but the ground state trimer does not.  
The excited trimer disappears through the 3-atom threshold
when $\lambda$ is decreased to about $0.97$, 
and it disappears through the atom-dimer threshold 
when $\lambda$ is increased to about 1.1.
The ground state trimer disappears through the 3-atom threshold
when $\lambda$ is about $0.9$. 
However, as $\lambda$ is increased above 1, its binding energy
relative to the atom-dimer threshold continues to increase. 
Thus the ground state $^4$He trimer does not qualify as an Efimov state 
by the traditional definition.

The traditional definition of an Efimov state described above is
not natural from the point of view of universality.
The essence of universality concerns the behavior of a system 
when the scattering length becomes increasingly large.
The focus of the traditional definition is on the endpoints 
of the binding energy curve in Fig.~\ref{fig:efistar},
which concerns the behavior of the system 
as the scattering length decreases in magnitude.
The problem is that the rescaling of the potential can move 
the system outside the universality region defined by $|a| \gg r_s$
before the trimer reaches the atom-dimer threshold.
We therefore propose a definition of an Efimov state
that is more natural from the universality perspective.
A trimer is defined to be an Efimov state if a deformation 
that tunes the scattering length to $\pm \infty$ moves its 
binding energy along the universal curve illustrated in 
Fig.~\ref{fig:efistar}.  The focus of this definition is on the 
resonant limit where the binding energy crosses the $1/a = 0$ axis.
In particular, the binding energy at this point should be larger 
than that of the next shallowest trimer by about a factor of 515.
In the case of $^4$He atoms, the resonant limit can be reached by rescaling
the 2-body potential by a factor $\lambda \approx 0.97$ \cite{Esry96a}.
At this point, the binding energy of the ground state trimer 
is larger than that of the excited trimer by about a factor of 570.
The closeness of this ratio to the asymptotic value 515
supports the hypothesis that the ground state $^4$He trimer
is an Efimov state. 

In order to apply the universal predictions for low-energy 3-body
observables to the case of $^4$He atoms,
we need a 2-body input and a 3-body input  
to determine the parameters $a$ and $\kappa_*$.  
The scattering length $a = 188.99 \, a_0$ for the TTY potential 
can be taken as the 2-body input.
An alternative 2-body input is the dimer binding energy: $E_2=1.31$ mK. 
A scattering length $a_D$ can be determined by identifying 
$E_2$ with the universal binding energy of the
shallow dimer: $E_2 = \hbar^2/(ma_D^2)$.
The result for the TTY potential is $a_D = 181.79 \, a_0$.
The 3.8\% difference between $a$ and $a_D$ is a measure of
how close the system is to the scaling limit.
To minimize errors associated with the
deviations of the system from the scaling limit, 
it is best to take the shallowest 3-body
binding energy available as the input for determining $\kappa_*$.  
In the case of $^4$He atoms, this is the binding energy 
$E_3^{(1)}$ of the excited trimer.
Experience has shown that the universal predictions are considerably 
more accurate if $E_2$ and $E_3^{(1)}$ are used as the inputs instead 
of $a$ and $E_3^{(1)}$ \cite{BH02}.

We proceed to consider the universal predictions for the 
trimer binding energies. Having identified $E_3^{(1)}$ 
with the universal trimer binding energy $E_T^{(1)}$, 
we can use Efimov's binding energy equation (\ref{B3-Efimov}) 
with $n_* = 1$ to calculate $\kappa_*$ 
up to multiplicative factors of $e^{\pi/s_0} \approx 22.7$
\cite{BH02}.  The result is 
$\kappa_* = 0.00215 \,a_0^{-1}$ or $\kappa_* = 0.00232 \,a_0^{-1}$, 
depending on whether $E_2$ or $a$ is used as the 2-body input.
The intuitive interpretation of $\kappa_*$ is that if a parameter 
in the short-distance potential is adjusted  to tune $a$ to $+\infty$, 
the binding energy $E_3^{(1)}$ 
should approach a limiting value of approximately $\hbar^2 \kappa_*^2/m$,
which is 0.201 mK or 0.233 mK depending on whether the 2-body input
is $E_2$ or $a$.

%%%%%%%%%%%%%%%%%%%%%%%%%%%%%%%%%%%%%%%%%%%%%%%%%%
\begin{table}[htb]
\begin{tabular}{l|cc|cccc}
              & $a$ & $a_D$ & $E_3^{(1)}$ & $E_3^{(0)}$ & $E_3^{(-1)}$    \\ 
\hline 
TTY potential & 100.0 &  96.2 &  2.28   &    126.4  &         --         \\
universality  &       & input &  input  &    129.1   & $5.38 \times 10^4$ \\
universality  & input &       &  input  &    146.4   & $6.23 \times 10^4$ \\
\hline
\end{tabular}
\vspace*{0.3cm}
\caption{
Binding energies $E_3^{(n)}$ of the $^4$He trimers
for the TTY potential (row 1)
compared to the universality predictions using as the inputs
either $E_2$ and $E_3^{(1)}$  (row 2) or $a$ and $E_3^{(1)}$  (row 3). 
Lengths are given in \AA \ and energies are given in mK.
The trimer binding energies for the TTY potential  
are from Ref.~\cite{RY00}. (Note that $\hbar^2/m=12.1194$ K\AA$^2$
for $^4$He atoms.)}
\label{tab:He1}
\end{table}
%%%%%%%%%%%%%%%%%%%%%%%%%%%%%%%%%%%%%%%%%%%%%%%%%%

Once $\kappa_*$ has been calculated, we can solve Eq.~(\ref{B3-Efimov}) 
for the binding energies of the deeper Efimov states.  
The prediction for the next two binding energies are shown in 
Table~\ref{tab:He1}.  The prediction for $E_3^{(0)}$
differs from the binding energy of the ground-state trimer by 2.6\% or 16.4\%,
depending on whether $E_2$ or $a$ is taken as the 2-body input.  
The expected order of magnitude of the error is the larger of
$\ell_{\rm vdW} / a = 5.4$\%  
and $(E_3^{(0)} / E_{\rm vdW} )^{1/2} \approx 50$\%. 
The errors are much smaller than $(E_3^{(0)} / E_{\rm vdW} )^{1/2}$, 
suggesting that the scaling limit is more robust than one might 
naively expect.
Efimov's equation (\ref{B3-Efimov}) also predicts infinitely many 
deeper 3-body bound states.  
The predictions for the binding energy $E_3^{(-1)}$ of the next deepest 
state are given in Table~\ref{tab:He1}.
The predictions are more than two orders of magnitude larger than the
van der Waals $E_{\rm vdW} \approx 420$~mK.
We conclude that this state and all the deeper bound states are artifacts 
of the scaling limit. 

Using the above values of $\kappa_*$ for the TTY potential,
we can immediately predict the atom-dimer scattering length $a_{AD}$.
If $a$ is used as the 2-body input, we find $a_{AD} \approx 0.94\, a$,
corresponding to $a_{AD}\approx 178\, a_0$.
If $E_2$ is used as the 2-body input,
we find $a_{AD} \approx 1.19\, a_D$, corresponding to  
$a_{AD}\approx 216\, a_0$.
These values are in reasonable agreement with the
calculation of Ref.~\cite{MSSK01}, which gave $a_{AD} = 248(10) \, a_0$.  
Since $r_s/ a = 7.3$\% for the TTY potential, 
much of the remaining discrepancy can perhaps be attributed 
to effective-range corrections.

Universality can also be used to predict the 3-body recombination rate 
constant for $^4$He atoms interacting through the TTY potential.  
The prediction for $\alpha_{\rm shallow}$ is $2.9 \, \hbar a_D^4/m$
or $6.9 \, \hbar a^4/m$, depending on whether 
$E_2$ or $a$ is used as the 2-body input.  
In either case, the coefficient of $\hbar a^4/m$ is much smaller than
the maximum possible value 67.1.  Thus $^4$He atoms are fortuitously 
close to a combination of $a$ and $\kappa_*$ for which 
$\alpha_{\rm shallow}$ is zero.  The 3-body recombination rate constant
has not yet been calculated for the TTY potential.
It has been calculated for the HFD-B3-FCI1 potential \cite{AJM95}.
The result is $\alpha_{\rm shallow} = 12 \times 10^{-29}$ cm$^6$/s 
\cite{SEGB02}, which compares well with the  universal prediction 
$9 \times 10^{-29}$ cm$^6$/s obtained using $E_2$ and $E_3^{(1)}$ as the
inputs \cite{bigrev}.

%%%%%%%%%%%%%%%%%%%%%%%%%%%%%%%%%%%%%%%%%%%%%%%%%%
%    Alkali atoms
%%%%%%%%%%%%%%%%%%%%%%%%%%%%%%%%%%%%%%%%%%%%%%%%%%

\subsection{Alkali atoms near a Feshbach resonance}
        \label{sec:deep-alkali}

Alkali atoms near a Feshbach resonance provide a unique window on 
Efimov physics, because the scattering length can be tuned 
experimentally.  Inelastic loss rates have proved to be an 
especially powerful probe of 3-body processes in these systems.
In this subsection, we discuss some key 
experiments on 3-body losses for
$^{23}$Na, $^{85}$Rb, $^{87}$Rb, and $^{133}$Cs atoms 
near Feshbach resonances and compare them with the predictions
from universality.

The first observation of the enhancement of inelastic losses 
near a Feshbach resonance was by the MIT group~\cite{Inouye98}.
They created Bose-Einstein condensate of $^{23}$Na atoms in the 
$|1,+1 \rangle$ hyperfine state and used a magnetic field 
to adjust the scattering length. 
They observed enhanced losses 
near the Feshbach resonances at 907 G and 853 G.
Since the $|1,+1 \rangle$ state is the lowest hyperfine state,
the inelastic losses come primarily from 3-body recombination.
The 3-body losses near the Feshbach resonance at 907 G
were studied systematically by the MIT group \cite{Stenger99}. 
They also studied 3-body
losses in a Bose-Einstein condensate of $^{23}$Na atoms in the 
$|1,-1 \rangle$ hyperfine state near a Feshbach resonance at 1195 G. 
The off-resonant scattering length in this region of high 
magnetic field is approximately 52 $a_0$, which is significantly 
smaller than the van der Waals scale 
$\ell_{\rm vdW} \approx 90 \, a_0$.
Near the Feshbach resonances at 907 G, the authors of 
Ref.~\cite{Stenger99} were able to increase the scattering length 
by about a factor of 5 over the off-resonant value.
However this it still not very large compared to the $\ell_{\rm vdW}$,
so the universal theory does not apply.
A quantitative description of the data on 3-body recombination 
near the 907 G resonance has been given using a scattering model 
for a Feshbach resonance with zero off-resonant scattering length 
\cite{Petrov04}.

%%%%%%%%%%%%%%%%%%%%%%%%%%%%%%%%%%%%%%%%%%%%%%%%%%
\begin{figure}[htb]
\bigskip
\centerline{\includegraphics*[width=8.5cm,angle=0]{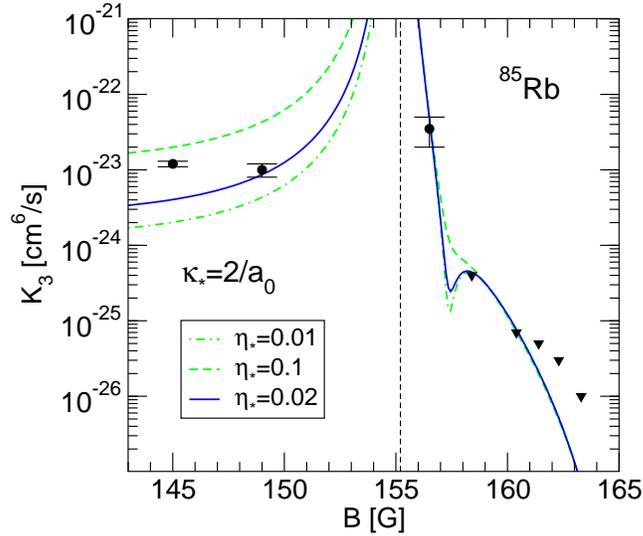}}
\medskip
\caption
{The 3-body loss coefficient $K_3$ in $^{85}$Rb as a function of the
magnetic field near the Feshbach  resonance at $B=155$ G. 
The dots are the measured values of $K_3$ from Ref.~\cite{Roberts00}, 
while the triangles indicate upper bounds on $K_3$.
The curves are for $\kappa_* a_0=2$ and several values of $\eta_*$.
}
\label{fig:85Rb}
\end{figure}
%%%%%%%%%%%%%%%%%%%%%%%%%%%%%%%%%%%%%%%%%%%%%%%%%%

The inelastic collision rate for ultracold $^{85}$Rb atoms in the 
$|2,-2\rangle$ hyperfine state near the Feshbach resonance at 155 G
has been studied by the JILA group \cite{Roberts00}.
By exploiting the different dependences on the number density, 
they were able to separate the contributions from 2-body 
and 3-body processes.  They measured the 2-body and 3-body loss 
coefficients as a function of the magnetic field $B$ from 110 G to 150 G.  
At some values of $B$, they were only able to obtain an upper bound 
on the 3-body loss coefficient $K_3 = 3\alpha$. 
Their results for $K_3$ near the Feshbach resonance at 155 G
are shown in Fig.~\ref{fig:85Rb}. 
The universal results for the 3-body recombination rate at threshold 
are given by Eq.~(\ref{alpha-deep:a<0}) for $a<0$ 
and they can be approximated by the sum of Eqs.~(\ref{alpha_sh:deep})
and (\ref{alpha-deep:a>0}) for $a>0$.
A precise determination of the Efimov parameters  $\kappa_*$ and $\eta_*$
is not possible with the data in Fig.~\ref{fig:85Rb}.  
However as shown in Fig.~\ref{fig:85Rb},
the data can be described reasonably well by the universal 
formulas with the values $\kappa_* a_0=2$ and $\eta_*=0.02$.
The curves for different values of  $\eta_*$ illustrate how the minima
in the recombination rate for $a>0$ are filled by recombination into
deep bound states.

The 3-body losses in ultracold $^{87}$Rb were studied 
by the Garching group \cite{Marte02}.
Roughly 90\% of the atoms were in the $|1,+1\rangle$ hyperfine state.
Most of the remaining atoms were in the $|1,0\rangle$ state,
but there were also some atoms in the $|1,-1\rangle$ state.
By monitoring the atom loss, the authors observed more than 40 
Feshbach resonances for various combinations of hyperfine states 
of $^{87}$Rb at magnetic fields ranging from 300 G to 1300 G.
Measurements of the resonances were used to deduce an improved atomic
potential for $^{87}$Rb. 
Away from the Feshbach resonances, the scattering
lengths for $^{87}$Rb have natural values comparable to the 
van der Waals scale $\ell_{\rm vdW} \approx 165 \, a_0$.
The 3-body loss rate  $K_3 = 3\alpha$ in the vicinity 
of the resonance at 1007 G 
was measured as a function of the magnetic field. 
Away from the resonance, $K_3$ was measured 
to be $3.2 (1.6) \times 10^{-29}$ cm$^6$/s.
Near the resonance, the rate constant was observed to increase 
by as much as a factor of 300.

The 3-body recombination rate in a Bose-Einstein condensate 
of $^{87}$Rb atoms in the $|1,+1\rangle$ hyperfine state has been 
measured just below the Feshbach resonance at 1007 G by 
the Garching group and by the Oxford group \cite{Smirne06}.
The values of the magnetic fields ranged from about 1 G below 
the resonance, 
where the scattering length has a natural value
comparable to the van der Waals scale 
$\ell_{\rm vdW} \approx 165 \, a_0$, to about 0.03 G below the
resonance, where the scattering length is 
about a factor 5 larger than $\ell_{\rm vdW}$.
Only the last few data points are in the universal region.
The measured rate constant $K_3=3\alpha$ appears to scale as a smaller power 
of $a$ than the scaling prediction $a^4$.
The measurements of $K_3$ agree reasonably well
with the results of exact solutions 
to the 3-body Schr{\"o}dinger equation for a coupled-channel 
model \cite{Smirne06}.  These calculations predict a local minimum of $K_3$
that can be attributed to Efimov physics at a magnetic field 
about 0.015 G below the resonance, just
outside the range of the experiments.
The atom-dimer relaxation rate for the coupled-channel 
model was also calculated in Ref.~\cite{Smirne06}.
The model predicts a resonance in the relaxation rate less than 0.1 G 
below the resonance.

The 3-body recombination rate for an ultracold gas of $^{133}$Cs atoms 
in the $|3,+3\rangle$ hyperfine state was measured as a function of the
magnetic field by the Innsbruck group \cite{Weber03}.
The 3-body recombination loss rate $K_3 = 3 \alpha$ is very small near 17 G, 
where the scattering length goes through a zero that comes from the 
interplay between a broad Feshbach resonance 
and the large off-resonant scattering length for $^{133}$Cs.
For magnetic fields between 50 G and 150 G,
the measurements confirmed the $a^4$ power-law 
dependence of $K_3$ predicted by scaling.  
In this region of the magnetic field,
the scattering length increased from about 1000 $a_0$ 
to about 1600 $a_0$.  
Since the scattering length increased only by
a factor of 1.6, it was not possible to observe any logarithmic 
variations of the coefficient of $a^4$.

%%%%%%%%%%%%%%%%%%%%%%%%%%%%%%%%%%%%%%%%%%%%%%%%%%
\begin{figure}[htb]
\bigskip
\centerline{\includegraphics*[width=8.5cm,angle=0]{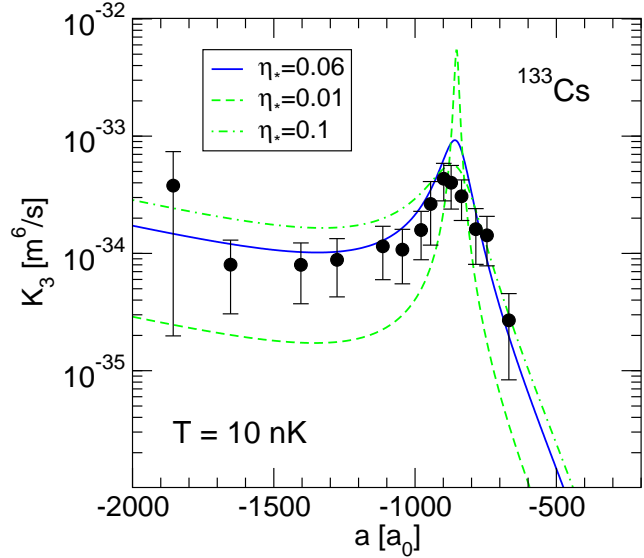}}
\medskip
\caption
{The 3-body loss coefficient $K_3$ in $^{133}$Cs for negative values of the 
scattering length near the Efimov resonance at $a\approx -850\,a_0$. 
The data points are for $T=10$ nK and are taken from Ref.~\cite{Kraemer06}.
The curves are for $\kappa_* = 0.945/a_0$ and
three different values of $\eta_*$.
}
\label{fig:133Cs}
\end{figure}
%%%%%%%%%%%%%%%%%%%%%%%%%%%%%%%%%%%%%%%%%%%%%%%%%%

The Innsbruck group extended their measurement of the 
3-body recombination loss rate  $K_3 = 3 \alpha$ 
for $^{133}$Cs atoms in the $|3,+3\rangle$ hyperfine state 
to values of the magnetic field below 17 G 
where the scattering length is negative~\cite{Kraemer06}.
The negative scattering length could be made as large as $-2500 \, a_0$.
They observed a resonant enhancement of $K_3$ for $a$ near $-800\,a_0$.
The enhancement can be explained by the presence of an Efimov state 
near the 3-atom threshold.
At the lowest temperature they were able to reach, which was 10 nK,
the shape of the resonance could be fit reasonably well using the 
universal result in Eq.~(\ref{alpha-deep:a<0}), which has a peak near $a_*$.  
The best fit values of the Efimov parameters are $a_*' = -850(20) \, a_0$
and $\eta_* = 0.06(1)$.
In Fig.~\ref{fig:133Cs}, we compare the data of Ref.~\cite{Kraemer06}
with the universal result.  The parameters $a_*'=-850 \, a_0$
(which corresponds to $\kappa_*=0.945 \,a_0^{-1}$) 
and $\eta_*=0.06$ give a good fit to the data.
Both Efimov parameters are well determined by the data: 
$\kappa_*$ by the position of the Efimov resonance 
and $\eta_*$ by the height and width of the peak.
Since the van der Waals length scale for $^{133}$Cs
is $\ell_{\rm vdW} \approx 200\,a_0$, the resonance is 
reasonably deep into the universal region.
Thus these results from Innsbruck group seem to be the first 
experimental evidence for the existence of Efimov states~\cite{Kraemer06}.

%%%%%%%%%%%%%%%%%%%%%%%%%%%%%%%%%%%%%%%%%%%%%%%%%%
\begin{figure}[htb]
\bigskip
\centerline{\includegraphics*[width=8.5cm,angle=0]{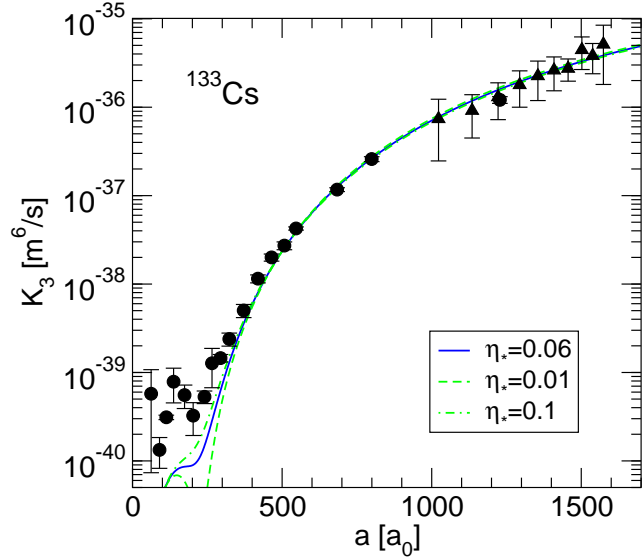}}
\medskip
\caption
{The 3-body loss coefficient $K_3$ in $^{133}$Cs for positive values of the 
scattering length.  The data points are for $T = 200$ nK (dots)
and for temperatures between 250 nK and 400 nK (triangles) and they are 
taken from Ref.~\cite{Kraemer06}.  The curves are for 
$\kappa_* = 0.707/a_0$ and three different values of $\eta_*$.
}
\label{fig:133Csp}
\end{figure}
%%%%%%%%%%%%%%%%%%%%%%%%%%%%%%%%%%%%%%%%%%%%%%%%%%

The Innsbruck group also carried out a more
careful analyisis of 3-body recombination at magnetic fields 
above 170 G, where $a$ is positive \cite{Kraemer06}.  
They measured the loss rate
$K_3$ with higher accuracy and at more values of the magnetic field.
They observed a local minimum in the atom loss rate for 
$a$ near $210 \, a_0$.  The dependence of the loss rate constant $K_3$ 
on the scattering length can be fit reasonably well with the universal 
formula given by the sum of Eqs.~(\ref{alpha_sh:deep})
and (\ref{alpha-deep:a>0}).  The value of  $\eta_*$ is not well 
determined by the data
and the fit yields only the upper bound $\eta_* < 0.2$.  
The best fit for the value 
of $a$ at which the coefficient of $a^4$ is maximal is
$a_+ = 1060 (70) \, a_0$.  This corresponds to $a_* \approx 1170 \, a_0$.
The positive scattering length data 
from Ref.~\cite{Kraemer06} are shown in Fig.~\ref{fig:133Csp}.
The Efimov parameter $\kappa_*=0.707 \, a_0^{-1}$
corresponding to  $a_* = 1170 \, a_0$
gives a good fit to the data above $a\approx 400\,a_0$.
If the scattering length could be increased past $\infty$
to a region where $a$ is large and negative, universality predicts 
that there should be an Efimov state at the 3-atom threshold when 
$a$ is approximately $- 1100 \, a_0$.  This is not far 
(at least on a log scale) from the position $- 850 \, a_0$ where the 
resonance was observed.  However the region 
of large negative $a$ in which the resonance is observed
is separated from the region of large positive $a$ by a region 
where $a$ passes through zero and universal theory does not apply.
As a consequence, the Efimov parameters for the two regions 
need not be equal. 
One note of caution in applying the universal expression
for the 3-body recombination rate at threshold is that  
the positive scattering length data in Fig.~\ref{fig:133Csp}
was obtained at temperatures that ranged from 200 nK to 400 nK, 
which is much larger than the temperature 10 nK for the negative 
scattering length data in Fig.~\ref{fig:133Cs}.
The temperature for the positive scattering length data
may not be low enough to use the universal 
expression for the 3-body recombination rate at threshold.

The Innsbruck group has also observed
inelastic losses of shallow dimers composed of $^{133}$Cs atoms 
in the $|3,+3\rangle$ hyperfine state \cite{Chin05}.
The fraction of dimers that are lost remains roughly constant for 
magnetic fields in the range from between 14 G and 19.8 G.
However, the fraction increases rapidly toward 1 as the magnetic 
field increases further toward a Feshbach resonance at 19.84 G.
In this experiment, resonant enhancements of the inelastic loss rate 
were also observed at lower values of the magnetic field.
They can be attributed to effects of a $^{133}$Cs tetramer near the 
threshold for two $^{133}$Cs dimers.  In this region of the magnetic 
field, the dimers are deeply bound, so these results have 
nothing to do with universality.

Very recently, the Innsbruck group has created mixtures
of $^{133}$Cs atoms in the $|3,+3\rangle$ hyperfine state
and shallow dimers composed of those atoms \cite{FB18Santos}. 
They have observed a resonance enhancement in the
inelastic atom-dimer collision rate near $a \approx 400 \ a_0$ 
\cite{FB18Santos}. 
The enhancement can be explained by the presence of an Efimov state 
near the atom-dimer threshold.  They have measured the rate constant
$\beta$ defined by Eq.~(\ref{dn-deact}) as a function of the 
magnetic field at a temperature near 250 nK.  
The universal predictions for $\beta$ as a function of $T$ have been
calculated for temperatures small compared to the binding 
energy of the shallow dimer. It might be possible 
to use these measurements to determine the Efimov parameters 
$a_*$ and $\eta_*$ accurately for this region of
large positive scattering length \cite{BH06}.

%\newpage 

%%%%%%%%%%%%%%%%%%%%%%%%%%%%%%%%%%%%%%%%%%%%%%%%%%
%
%     UNIVERSALITY IN OTHER THREE-BODY SYSTEMS
%
%%%%%%%%%%%%%%%%%%%%%%%%%%%%%%%%%%%%%%%%%%%%%%%%%%

\section{Efimov Physics in Other Three-body Systems}
        \label{sec:beyond}

%%%%%%%%%%%%%%%%%%%%%%%%%%%%%%%%%%%%%%%%%%%%%%%%%%
%   Unequal scattering lengths
%%%%%%%%%%%%%%%%%%%%%%%%%%%%%%%%%%%%%%%%%%%%%%%%%%

Very few universal results have been calculated for 3-body systems 
other than identical bosons.  They are summarized in Ref.~\cite{bigrev}.
In this section, we consider only the basic question of whether the 
Efimov effect occurs in the 3-body system.  
This question can be answered by determining
the channel eigenvalue $\lambda_0(R)$ for the lowest hyperspherical potential
in the scaling limit.  The Efimov effect occurs if $\lambda_0(R)$ is negative 
at $R = 0$.  If $\lambda_0(0) = - s_0^2$, the discrete scaling factor
for Efimov physics is $e^{\pi/s_0}$.

\subsection{Unequal scattering lengths}
       \label{sec:unequal}

There are 3-body systems in atomic physics in which the three atoms 
all have the same mass, 
but the three pairs of atoms need not all have the same scattering lengths.
For example, different hyperfine spin states of the same atom have the same 
mass, but the scattering lengths $a_{ij}$ can be different for each pair
$ij$ of hyperfine states.   As another example, different isotopes of a heavy 
atom have nearly the same masses.  
It is therefore worthwhile to consider the universal 
behavior of systems with equal masses and with scattering lengths 
that are large but not necessarily equal.

The Efimov effect in general 3-body 
systems was first discussed by Amado and Noble \cite{AN72}
and by Efimov \cite{Efimov72,Efimov73}.  
A summary of their results is as follows. If only one of the 
three scattering lengths is large, the Efimov effect does not occur.
If two of the scattering lengths are large,
the Efimov effect occurs with a discrete scaling factor of 1986.1
unless two of the three particles are identical fermions,
in which case the Efimov effect does not occur.
If all three scattering lengths are large,
the Efimov effect occurs with a discrete scaling factor of 22.7.
For a derivation of these results, see Ref.~\cite{bigrev}.

If the atoms are distinguishable but related by an internal symmetry,
the condition for the Efimov effect is modified.
An example is the spin states belonging to a hyperfine multiplet 
at zero magnetic field, which are related by a spin symmetry.
A general treatment of the effects of an internal symmetry 
was given by Bulgac and Efimov in Ref.~\cite{BE75}.

%%%%%%%%%%%%%%%%%%%%%%%%%%%%%%%%%%%%%%%%%%%%%%%%%%
%    Unequal masses
%%%%%%%%%%%%%%%%%%%%%%%%%%%%%%%%%%%%%%%%%%%%%%%%%%

\subsection{Unequal masses}
\label{sec:mass}

In the 2-body sector, the universal results for particles of unequal 
masses are only a little more complicated than those for the equal-mass
case in Sections~\ref{sec:uni2AA} and \ref{sec:uni2D}.
Let the atoms 1 and 2 have a large scattering length $a_{12}$
and unequal masses $m_1$ and $m_2$.  
If $a_{12}$ is large and positive, the atoms 1 and 2 form 
a shallow dimer with binding energy
%----------------------
\begin{eqnarray}
E_D = 
{\hbar^2 \over 2 m_{12} a_{12}^2} \,,
\end{eqnarray}
%----------------------
%[E,H]
where $m_{12} = m_1 m_2/(m_1 + m_2)$ is the reduced mass.

In the general 3-body system, the three masses can be unequal 
and any combination of the three scattering lengths can be large.
The Efimov effect in general 3-body 
systems was first discussed by Amado and Noble \cite{AN72}
and by Efimov \cite{Efimov72,Efimov73}.  The special case in which two of the 
three particles have the same mass was also discussed by
Ovchinnikov and Sigal \cite{OS79}. The conditions for the
existence of the Efimov effect and the value of the discrete scaling factor 
depend on the ratios of the masses.
We first summarize the results for the extreme cases
in which two masses are equal and the third mass 
is either much larger or much smaller.
In the case of two heavy particles and one light particle, 
the Efimov effect occurs provided the heavy-light scattering length is large.
The Efimov effect can be understood intuitively in this case
by using the Born-Oppenheimer approximation \cite{FRS79}. 
In the case of one heavy particle and two light particles, the Efimov effect 
occurs only if all three pairs of particles have large scattering lengths.

%%%%%%%%%%%%%%%%%%%%%%%%%%%%%%%%%%%%%%%%%%%%%%%%%%
\begin{figure}[htb]
\medskip
\centerline{\includegraphics*[width=8.5cm,angle=0]{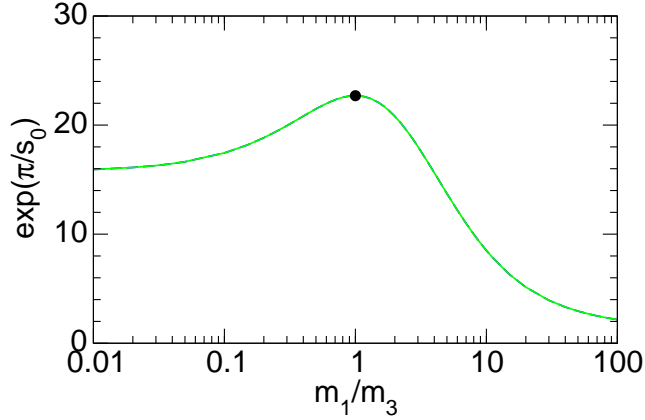}}
\medskip
\caption{
Discrete scaling factor $e^{\pi/s_0}$ 
for two particles of equal mass $m_1 = m_2$ as a function of the 
mass ratio $m_1/m_3$ for the case 
in which all three pairs have large scattering lengths.
The particles 1 and 2 can be either identical bosons or 
distinguishable. The dot indicates the case of three identical bosons.
}
\label{fig:dsf2}
\end{figure}
%%%%%%%%%%%%%%%%%%%%%%%%%%%%%%%%%%%%%%%%%%%%%%%%%%

We next consider a more general case in which all three pairs 
of atoms have a large scattering length. This excludes the possibility 
of any pair of particles being identical fermions. 
The condition for the Efimov effect 
does not depend on the values of the scattering lengths 
as long as their absolute values are large compared to the range. 
The Efimov effect occurs for any values of the masses.
The discrete scaling factor is largest if all three masses are equal.
It has the same value 
$e^{\pi / s_0} \approx 22.7$ as for three identical bosons.  
In the case of two equal-mass particles,
the discrete scaling factor is the same whether
the equal-mass particles are identical bosons or distinguishable.
The discrete scaling factor for $m_1 = m_2$
is shown as a function of the mass ratio 
$m_1/m_3 = m_2/m_3$ in Fig.~\ref{fig:dsf2}.
In the limit $m_1 = m_2 \ll m_3$, 
the discrete scaling factor approaches $15.7$.
In the limit $m_1 = m_2 \gg m_3$, it approaches 1.

%%%%%%%%%%%%%%%%%%%%%%%%%%%%%%%%%%%%%%%%%%%%%%%%%%
\begin{figure}[htb]
\medskip
\centerline{\includegraphics*[width=8.5cm,angle=0]{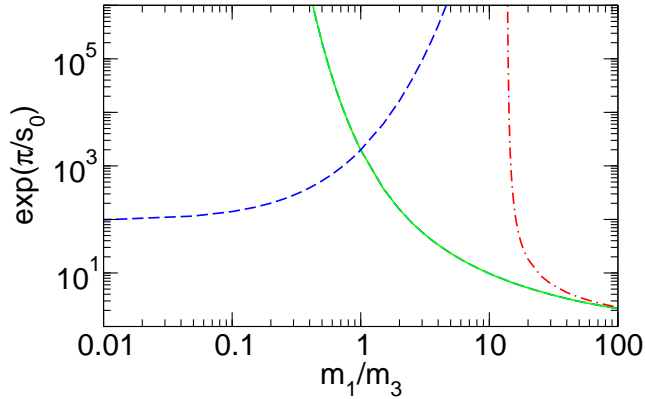}}
\medskip
\caption{
Discrete scaling factor $e^{\pi/s_0}$ for two particles of equal mass 
$m_1 = m_2$ as a function of the mass ratio 
$m_1/m_3$ for the cases in which two pairs have large scattering lengths.
If $a_{23}$ and $a_{31}$ are large, particles 1 and 2 
can be either identical bosons or distinguishable particles (solid line)
or else identical fermions (dash-dotted line).
If $a_{12}$ and $a_{31}$ (or $a_{12}$ and $a_{23}$) are large,
particles 1 and 2 must be distinguishable particles (dashed line).
}
\label{fig:dsf1}
\end{figure}
%%%%%%%%%%%%%%%%%%%%%%%%%%%%%%%%%%%%%%%%%%%%%%%%%%

We now consider the more general case in which only two pairs of atoms 
have large scattering lengths.  For simplicity,
we consider only the special cases in which particles 1 and 2 have 
the same masses $m_1 = m_2$.  There are three cases to consider.
The discrete scaling factors are shown in Fig.~\ref{fig:dsf1}
as a function of the mass ratio $m_1/m_3$ for all three cases.
If the large scattering lengths are $a_{23}$ and $a_{31}$ 
and if particles 1 and 2 are either identical bosons or distinguishable,
the discrete scaling factor $e^{\pi/s_0}$ decreases monotonically 
from $\infty$ to 1986.
If the large scattering lengths are $a_{23}$ and $a_{31}$ 
and if particles 1 and 2 are identical fermions, 
the Efimov effect occurs only if $m_1/m_3$ exceeds the critical value 13.607.
The discrete scaling factor decreases monotonically from 
$\infty$ to 1 as $m_1/m_3$ increases from the critical 
value to $\infty$.
If the large scattering lengths are $a_{12}$ and $a_{31}$ 
(or $a_{12}$ and $a_{23}$), particles 1 and 2 must be distinguishable.
The discrete scaling factor increases monotonically from 94.36 to 1986.1
to $\infty$ as $m_1/m_3$ increases from 0 to 1 to $\infty$.

%\newpage

%%%%%%%%%%%%%%%%%%%%%%%%%%%%%%%%%%%%%%%%%%%%%%%%%%
%
%     BEYOND THE SCALING LIMIT
%
%%%%%%%%%%%%%%%%%%%%%%%%%%%%%%%%%%%%%%%%%%%%%%%%%%

\section{Extensions and Outlook}
        \label{sec:beyondsl}

In this section, we discuss extensions of 
universality beyond the 3-body problem in the scaling limit: 
effective-range corrections, microscopic 
models with a large scattering length, and the 4-body problem.
We end this review with an outlook on the study of Efimov physics.

\subsection{Effective range corrections}
        \label{sec:erc}

The scaling limit of a system with a large scattering length 
is the limit in which the range of the interaction is taken to zero.
There are corrections to the scaling limit from the fact that
the range for any physical system is nonzero. These corrections 
can be calculated as an expansion in powers of $\ell/a$,
where $\ell$ is the natural low-energy length scale.
The most important corrections come from the effective range
$r_s$, which is defined by the expansion of the S-wave phase 
shift in Eq.~(\ref{kcot}).  Examples of effective range corrections
for 2-body observables are given in Eqs.~(\ref{dsig:exp})
and (\ref{B2-exp}).

The most important correction to the scaling limit in the 3-body 
system also comes from the effective range $r_s$.  At first order 
in the expansion in $\ell/|a|$, the effective range 
is the only new parameter that enters.  Efimov has shown that 
the effective range leads to a correction term $r_s/R^3$
in the scale-invariant hyperspherical potential 
in Eq.~(\ref{Vch}) \cite{Efimov93}. 
He gave expressions for the first-order effective range corrections
to the binding energy of the Efimov trimer
and to the atom-dimer scattering length \cite{Efimov91}.
Moreover, he applied his results to the 3-nucleon system, calculating the
range corrections to the correlation between the triton binding energy and
the spin-doublet neutron-deuteron scattering length \cite{EfT85}
and to the spin-quartet neutron-deuteron scattering length
\cite{Efimov91}.

Here, we will give only a brief summary of the systematics of higher-order 
range corrections in channels where the Efimov effect occurs. 
For a more complete discussion, we refer the reader to our longer review 
\cite{bigrev}. Since the correction to the universal long-range 
potential is singular, the range corrections require renormalization.
Effective field theory provides a powerful framework to deal
with these naively divergent corrections.
The first-order effective range correction 
to the spin-doublet neutron-deuteron scattering phase
shifts was first calculated using this method in \cite{HM-01b}
and later confirmed using a different renormalization scheme 
\cite{Afnan:2003bs}.
The systematics of contributions beyond first order in $\ell/|a|$ was
studied by various authors 
\cite{BGHR03,Griesshammer:2005ga,Birse:2005pm,Platter:2006ev,Platter:2006ad}.
The first calculation of the triton channel of the 3-nucleon system
at second order in $\ell/|a|$ was carried out in Ref.~\cite{BGHR03}.
As suggested by naive dimensional analysis, 
a new 3-body parameter was included in this calculation 
at order $(\ell/|a|)^2$.
Recently, Platter and Phillips revisited this problem using a 
subtractive renormalization scheme.
They found that the scattering amplitudes could
be renormalized at order $(\ell/|a|)^2$ without the need of a 
new subtraction constant \cite{Platter:2006ev,Platter:2006ad}. 
According to their analysis, a new 3-body parameter
is not required until at least order $(\ell/|a|)^3$.
They applied their method to the system of $^4$He atoms and found 
good convergence of the expansion in $\ell/|a|$ \cite{Platter:2006ev}.

\subsection{Microsopic models}
        \label{sec:feshbach}

Since the low-energy behavior of particles with large scattering 
lengths is insensitive to the details of their interactions at 
short distances, the potential provides an inefficient encoding 
of the relevant physics.  The sensitivity to short distances 
enters primarily through the scattering length and other constants 
that describe low-energy scattering.  This motivates a more 
phenomenological approach in which the interactions are 
described by a {\it scattering model}, which can be specified by a 
parameterization of low-energy scattering amplitudes.  
The parameters of the scattering model can be treated as 
phenomenological parameters that can be tuned to reproduce 
the observed low-energy observables of the 2-body system.  
Scattering models will exhibit universal low-energy behavior 
as the scattering length $a$ is tuned to $\pm \infty$.  
Low-energy 3-body observables reduce in this limit 
to functions of $a$ and the Efimov parameters 
$\kappa_*$ and $\eta_*$, all of which are functions 
of the parameters of the scattering model.  

An example of a scattering model that may provide an accurate 
description of atoms near a Feshbach resonance is 
the {\it resonance model} \cite{KMCWH02}.  
This model provides a natural description of atoms 
that have a weak coupling to a closed channel 
in which there is a diatomic molecule extremely close to the 
atom-atom threshold.  The resonance model has three 
parameters that can be defined by specifying 
the S-wave phase shift $\delta_0(k)$: 
%----------------------
\begin{eqnarray}
k \cot \delta_0(k) = 
- \left [\lambda + \frac {g^2}{k^2 - \nu} \right ]^{-1} \,.
\label{kcot:RM}
\end{eqnarray}
%----------------------
%[E]
The scattering length and the effective range are
$a = \lambda - g^2/\nu$ and $r_s = -2(g - \nu \lambda/g)^2$,
respectively. Thus the effective range is negative definite. 
The approximation in Eq.~(\ref{Feshbach}) for the scattering length 
as a function of the magnetic field $B$ for an atom near a 
Feshbach resonance can be reproduced by taking $\lambda$ 
and $g$ to be independent of $B$ and $\nu$ to be linear in 
$B-B_{\rm res}$.  Thus $\lambda$ can be identified with the 
off-resonant scattering length, $\nu$ is proportional 
to the detuning energy of a molecule in a closed channel, 
and $g$ is the strength of the coupling to the closed channel.  
The resonance model 
will exhibit universal behavior as $\nu$ is tuned to zero.  
Three-body observables will approach universal functions of $a$,
$\kappa_*$, and $\eta_*$.  The Efimov parameter $\kappa_*$
must be a function of $\lambda$ 
and $g$, while $\eta_*$ must be a 
function of the dimensionless combination $\lambda g^2$.

Some 3-body observables have been calculated in the 
resonance model in the special case $\lambda = 0$ \cite{Petrov04}.  
The scattering length and the effective range are $a = -g^2/\nu$
and $r_s = -2/g^2$, respectively.  It is convenient to define a 
parameter $R_* = 1/g^2$, with dimension of length. The atom-dimer 
scattering length $a_{AD}$ and the 3-body recombination rate
into the shallow dimer $\alpha$ have been calculated in this model 
as functions of $a$ and $R_*$ \cite{Petrov04}.  For $a \gg R_*$, 
they have the universal behavior 
given in Eqs.~(\ref{a12-explicit}) and (\ref{alpha-analytic}).  
The 3-body parameter $\kappa_*$ was determined 
in Ref.~\cite{Petrov04} to be
%----------------------
\begin{eqnarray}
s_0 \ln (\kappa_*) = s_0 \ln (2.5 /R_*) \mod \pi \,.
\end{eqnarray}
%----------------------
%[E]
The cross-over to the universal behavior occurs when $a$ is 
comparable to $R_*$.  The first divergence of the atom-dimer 
scattering length occurs at $a = 0.45 \, R_*$.  This marks the 
emergence of the first Efimov state below the atom-dimer threshold.  
The second and higher divergences of $a_{AD}$ occur at values 
of $a$ that are well-approximated by the universal predictions: 
$a = (e^{\pi /s_0})^n \, 0.64 \, R_*$, $n = 1, 2, \dots$, 
where $e^{\pi/s_0} \approx 22.7$.  The first zero of the 
recombination rate constant occurs at $a = 3.3 \, R_*$.  
The second and higher zeroes occur at values of $a$ that are
well approximated by the universal predictions:
$a = (e^{\pi/ s_0})^n \, 2.9 \, R_*$, $n = 1, 2, \dots\;$.

\subsection{Four-body problem}
        \label{sec:fourbody}

In the 3-body problem, exact numerical solutions are facilitated by the 
Faddeev equations.
The generalization of the Faddeev equations to the $N$-body problem 
with $N \ge 4$ was given by Yakubovsky \cite{Yakubovsky:1966ue}.
An equivalent 
set of equations was given independently by Grassberger and Sandhas
\cite{GrSa67}. Due to the complexity of these equations, exact 
numerical solutions for $N=4$ have only recently been obtained.
In atomic physics, the only numerically exact $N$-atom calculations 
for $N \ge 4$ that we are aware of are for ground-state binding energies.
There are no such calculations for weakly-bound $N$-atom molecules
that might be governed by universality.

Testing universality in systems with four and more particles is a new
frontier. Even theoretically, the universal properties in the 4-body 
sector are not fully understood. 
An important issue in the 4-body system with a large scattering length
is how many parameters are required to describe the system in the scaling 
limit, that is, up to corrections that decrease like $\ell/|a|$ as 
$a \to \pm \infty$. In the case of identical bosons,
low-energy 4-body observables necessarily 
depend on the 2-body parameter $a$ and the 3-body parameter $\kappa_*$.
But are any additional 4-body parameters required?
There are theoretical arguments in support of both answers to this question. 
There is a renormalization argument for zero-range  2-body
potentials that indicates that an additional $N$-body parameter is required
to calculate $N$-body binding energies for all $N \ge 3$ \cite{AFG95}. 
On the other hand, a power-counting argument within the effective
field theory framework suggests that after adding the 3-body 
parameter $\kappa_*$, no additional parameters
are necessary to calculate $N$-body observables for $N \ge 4$  
\cite{Lepage-pc}. 

Platter et al. have recently studied the universal properties of the 
system of four identical bosons with short-range interactions in an
effective quantum mechanics approach \cite{Platter:2004qn}.
They constructed the effective interaction potential at leading order 
in the large scattering length and computed the 4-body binding energies 
using the Yakubovsky equations. They found that cutoff independence of the 
4-body binding energies does not require the introduction of a 4-body 
parameter at leading order in $\ell/|a|$. 
This suggests that 2-body and 3-body interactions are sufficient 
to renormalize the 4-body system. They applied their equations 
to $^4$He atoms and calculated the binding energies of the 
ground state and the excited state of the $^4$He tetramer. 
Using the binding energy $E_2$ of the $^4$He  dimer as the 2-body input 
and the binding energy $E_3^{(1)}$ of the excited state of the $^4$He trimer 
as the  3-body input, they found good agreement 
with results of Blume and Greene \cite{BlGr00} calculated
using an approximate numerical method that combines Monte Carlo methods
with the hyperspherical expansion.

Yamashita et al.~\cite{Yamashita06} recently claimed that for a
4-body system close to a Feshbach resonance, a new 4-body 
parameter will enter already at leading order in $\ell/|a|$. 
They motivated their claim from a model-space reduction of a
realistic 2-body interaction. In the framework of the 
renormalized zero-range model, they found a strong sensitivity of 
the 4-body ground state energy to a 4-body subtraction constant
in their equations. The results of Ref.~\cite{Platter:2004qn} 
for the special case of $^4$He atoms were also reproduced.
A drawback of their analysis is the focus on the deepest 4-body
state only.  It remains to be seen
whether their findings are universal or just artefacts of
their regularization of the renormalized zero-range model.

%%%%%%%%%%%%%%%%%%%%%%%%%%%%%%%%%%%%%%%%%%%%%%%%%%
%    OUTLOOK
%%%%%%%%%%%%%%%%%%%%%%%%%%%%%%%%%%%%%%%%%%%%%%%%%%

\subsection{Outlook}

Vitaly Efimov's discovery in 1970 of the Efimov effect in identical
bosons was like finding a nugget of a precious metal.  Other
physicists quickly verified the amazing properties of
this nugget and tried to find similar nuggets in other systems. 
Efimov went on to show that the nugget came
from a rich vein of ore corresponding to universal properties of 3-body
systems with a large scattering length. He pointed out that the 
Efimov effect was just a hint of a discrete scaling symmetry,
and showed how the discrete scaling symmetry could be exploited
to mine the ore.  However not many physical systems in which 
Efimov physics plays an important role were found. 
In those that were identified, such as $^4$He atoms,
it was difficult to study Efimov physics experimentally.
Efimov's mine was eventually closed due to lack of interest from other
physicists as well as the lack of a market for the ore. 

The new field of cold atom physics has changed the situation
dramatically. The development of the technology for trapping alkali atoms
and cooling them to ultralow temperatures has provided a long list of
3-body systems in which Efimov physics can be studied. Especially
important is the possibility of using Feshbach resonances to control the
scattering length and make it arbitrarily large. This development has
justified the reopening of Efimov's mine.

The theoretical study of Efimov physics is still in its infancy. 
In the case of 3 identical bosons, the rates for most of the basic 
scattering processes have been calculated only at threshold. 
They can be used to make universal predictions 
for cold atoms at sufficiently low temperatures. 
To determine the Efimov parameters $\kappa_*$ and $\xi_*$ accurately 
and to test the accuracy of the universal predictions, it is important 
to have universal predictions for rate constants as a function of 
temperature.  This requires the calculation of the rates for the basic 
scattering processes as functions of the collision energy. 
The only such calculation thus far is the S-ware atom-dimer phase shift 
at collision energies up to the dimer break-up threshold \cite{BH02}. 
For 3-body systems other than identical bosons that also exhibit 
the Efimov effect, the universal results for the basic
scattering processes have not even been calculated at threshold.

The calculation of range corrections is important for improving the
accuracy of universal predictions and also for quantifying their domain
of validity. Although methods for calculating range corrections have been
developed, they have not been implemented in a way that allows them to be
easily applied to experiment. The calculation of 3-body observables in
microscopic models for atoms near Feshbach resonance would also be
useful for determining the range of validity of the universal
predictions.

If the theoretical study of Efimov physics is still in its infancy, 
the experimental study of Efimov physics is a newborn. 
The beautiful results on $^{133}$Cs atoms from the Innsbruck group 
have provided the first glimpse of Efimov physics in cold atoms. 
These experiments have demonstrated that Efimov physics 
can have dramatic effects in alkali
atoms in spite of their many deep 2-body bound states. 
They inspire confidence that Efimov physics will also be observable 
in other cold atom systems.

The next milestone in the experimental study of Efimov physics in cold
atoms will be quantitative tests of the correlations between 3-body
observables predicted by universality. For example, if $a$ is large and
positive, universality gives correlations between atom-dimer scattering
and 3-body recombination. If a Feshbach resonance is used to tune the
scattering length through $\pm \infty$, universality gives correlations
between 3-body recombination at large positive $a$ on one side of the
resonance and 3-body recombination at large negative $a$ on 
the other side of the resonance.

The ultimate confirmation of Efimov physics would be the experimental
verification of the discrete scaling symmetry. This could be accomplished
by the observation of loss features in 3-body recombination or in 
atom-dimer scattering at two or more values of $a$ that differ by powers 
of the discrete scaling factor. For 3-body recombination with identical
bosons, the large size of the discrete scaling factor 22.7 and the $a^4$
scaling of the loss rate makes this difficult.
Atom-dimer scattering is more favorable in this regard because the loss
rate scales only as $a$. Verification of the discrete scaling symmetry
would be easier in a system consisting of two heavy atoms and a 
third light atom because the discrete scaling factor 
can be much smaller than that for identical bosons.
One particularly favorable possibility is the system consisting of 
two $^{133}$Cs atoms and a $^{6}$Li atom near a Feshbach resonance 
in the Cs-Li interaction.  This system has a discrete scaling factor 
of only 4.88 \cite{dinc06}.

The discovery of the Efimov effect in 1970 provided a hint of the
remarkable phenomena associated with Efimov physics 
that was waiting to be uncovered. 
For 35 years the study of Efimov physics was an almost
exclusively theoretical enterprise that could be characterized as
an exploratory mine shaft. The advent of cold atom physics 
made it inevitable that Efimov physics would also be studied 
experimentally. This field offers a long list of systems in which
Efimov physics could occur, 
as well as offering the possibility of tuning the
scattering length experimentally. An experimental mine shaft into 
Efimov physics has finally been opened by the Innsbruck experiment 
on $^{133}$Cs atoms.  This should stimulate 
the opening of mine shafts into Efimov physics in other 
cold atom systems.  Extensive experimental effort to extract
the ore and additional theoretical effort to refine the ore
will be required to reveal the full beauty of Efimov physics.

\section*{Acknowledgments}
E.B.~was supported by Department of Energy grants DE-FG02-91-ER4069
and DE-FG02-05ER15715. 

%\newpage  


\begin{thebibliography}{99}

\bibitem{Efimov70}
V.~Efimov,
% ``Energy Levels Arising from Resonant Two-body Forces
%        in a Three-body System,''
Phys.\ Lett.\ {\bf 33B}, 563 (1970).

\bibitem{Efimov71}
V.~Efimov,
% ``Weakly-bound States of Three Resonantly-interacting Particles,''
Sov.\ J.\ Nucl.\ Phys.\ {\bf 12}, 589 (1971)
[Yad.\ Fiz.\ {\bf 12}, 1080 (1970)].

\bibitem{Efimov79}
V.~Efimov,
% ``Low-energy Properties of Three Resonantly-interacting Particles,''
Sov.\ J.\ Nucl.\ Phys.\ {\bf 29}, 546 (1979)
[Yad.\ Fiz.\ {\bf 29}, 1058 (1979)].

\bibitem{STo96} 
W.~Sch\"ollkopf and J.P.~Toennies, 
% ``The nondestructive detection of the helium dimer and trimer,''
J.\ Chem.\ Phys.\  {\bf 104}, 1155 (1996).

\bibitem{Kraemer06}
T.~Kraemer, M.~Mark, P.~Waldburger, J.G.~Danzl, C.~Chin, B.~Engeser,
A.D.~Lange, K.~Pilch, A.~Jaakkola, H.-C.~N\"agerl, and R.~Grimm,
% "Evidence for Efimov Quantum States in an Ultracold Gas of Caesium 
% Atoms,"
Nature {\bf 440}, 315 (2006).

\bibitem{bigrev}
  E.~Braaten and H.-W.~Hammer,
%  ``Universality in Few-body Systems with Large Scattering Length,''
  Phys.\ Rept.\  {\bf 428}, 259 (2006).
%  [arXiv:cond-mat/0410417].
  %%CITATION = COND-MAT 0410417;%%

\bibitem{Schwinger47}
J.~Schwinger,
% ``A Variational Principle for Scattering Problems,''
Phys.\ Rev.\ {\bf 72}, 742 (1947).

\bibitem{AS91}
R.A.~Aziz and M.J.~Slaman, 
% ``An examination of ab initio results 
%        for the helium potential energy curve,'' 
J.\ Chem.\ Phys.\ {\bf 94}, 8047 (1991).

\bibitem{TTY95}
K.T.~Tang, J.P.~Toennies, and C.L.~Yiu, 
% ``Accurate Analytical He-He van der Waals Potential 
%         Based on Perturbation Theory,'' 
  Phys.\ Rev.\ Lett.\ {\bf 74}, 1546 (1995).

\bibitem{BEGKH02}
D.~Blume, B.D.~Esry, C.H.~Greene, N.N.~Klausen, and G.J.~Hanna,
% ``Formation of Atomic Tritium Clusters and Bose-Einstein Condensates,''
Phys.\ Rev.\ Lett.\ {\bf 89}, 163402 (2002). 
%        [arXiv:physics/0207002].

\bibitem{Feshbach62}
H.~Feshbach, 
% ``A unified theory of nuclear reactions. II,''
Ann.\ Phys.\ {\bf 19}, 287 (1962).

\bibitem{TMVS92}
E.~Tiesinga, A.J.~Moerdijk, B.J.~Verhaar, and H.T.C.~Stoof, 
% ``Conditions for Bose-Einstein condensation in magnetically trapped 
%         atomic cesium,''
Phys.\ Rev.\ A {\bf 46}, R1167 (1992).

\bibitem{TVS93}
E.~Tiesinga, B.J.~Verhaar, and H.T.C.~Stoof, 
% ``Threshold and resonance phenomena in ultracold ground-state collisions,''
Phys.\ Rev.\ A {\bf 47}, 4114 (1993). 

\bibitem{Inouye98}
S.~Inouye, M.R.~Andrews, J.~Stenger, H.-J.~Miesner, D.M.~Stamper-Kurn, 
        and W.~Ketterle,
% ``Observation of Feshbach Resonances in a Bose-Einstein condensate,''
Nature {\bf 392}, 151 (1998).

\bibitem{Stenger99}
J.~Stenger, S.~Inouye, M.R.~Andrews, H.-J.~Miesner, D.M.~Stamper-Kurn,
and W.~Ketterle,
% "Strongly Enhanced Inelastic Collisions in a Bose-Einstein Condensate
% near Feshbach Resonances,"
Phys.\ Rev.\ Lett.\ {\bf 82}, 2422 (1999).

\bibitem{Courteille98} 
Ph.~Courteille, R.S.~Freeland, D.J.~Heinzen, F.A.~van Abeelen,
        and B.J.~Verhaar,
% ``Observation of a Feshbach Resonance in Cold Atom Scattering,''
Phys.\ Rev.\ Lett.\ {\bf 81}, 69 (1998).

\bibitem{Roberts98} 
J.L.~Roberts, N.R.~Claussen, J.P.~Burke, Jr., C.H.~Greene, 
        E.A.~Cornell, and C.E.~Wieman,    
% ``Resonant Magnetic Field Control of Elastic Scattering in Cold 85Rb'',
Phys.\ Rev.\ Lett.\ {\bf 81}, 5109 (1998).

\bibitem{FTDA99}
T.~Frederico, L.~Tomio, A.~Delfino, and A.E.A.~Amorim,
% ``Scaling limit of weakly bound triatomic states,''
Phys.\ Rev.\ A {\bf 60}, R9 (1999).

\bibitem{MiF62}
R.A.~Minlos and L.D.~Faddeev, 
% ``On the point interaction for a three-particle system in quantum
% mechanics,''
Sov.\ Phys.\ Doklady {\bf 6}, 1072 (1962)
   [Doklady Akademii Nauk SSR {\bf 141}, 1335 (1961)].

\bibitem{MiF62b}
R.A.~Minlos and L.D.~Faddeev, 
% ``Comment on the Problem of Three Particles with Point Interactions,''
Sov.\ Phys.\ JETP {\bf 14}, 1315 (1962)
        [J.\ Exptl.\ Theoret.\ Phys.\ (U.S.S.R.) {\bf 41}, 1850 (1961)].

\bibitem{NFJG01}
E.~Nielsen, D.V.~Fedorov, A.S.~Jensen, and E.~Garrido,
% ``The Three-Body Problem with Short-Range Interactions,''
Phys.\ Rep.\ {\bf 347}, 373 (2001).

\bibitem{Delves60}
L.M.~Delves, 
% ``Tertiary and general-order collisions. II'',
Nucl.\ Phys.\ {\bf 20}, 275 (1960).

\bibitem{ZM88}
Z.~Zhen and J.~Macek, 
%``Loosely bound states of three particles,''
Phys.\ Rev.\ A {\bf 38}, 1193 (1988).

\bibitem{BHK-02a}
E.~Braaten, H.-W.~Hammer, and M.~Kusunoki,
% ``Universal Equation for Efimov States,''
Phys.\ Rev.\ A {\bf 67}, 022505 (2003).  
%        [arXiv:cond-mat/0201281].

\bibitem{Mohr03}
R.F.~Mohr, 
% ``Quantum Mechanical Three-Body Problem with Short-Range Interactions,''
Ph.D thesis, Ohio State University (2003) 
        [arXiv:nucl-th/0306086].

\bibitem{Jonsell06}
S.~Jonsell,
% ``Efimov states for systems with negative scattering lengths,''
Europhys.\ Lett.\ {\bf 76}, 8 (2006).

\bibitem{YFT06}
M.T.~Yamashita, T.~Frederico, and L.~Tomio,
% ``Three-boson recombination at ultralow temperatures,''
arXiv:cond-mat/0608542.

\bibitem{Sim81}
I.V.~Simenog and A.I.~Sitnichenko, 
%``Effect of long-range interaction in a three-body system
%        with short-range forces,''
Doklady Academy of Sciences of the Ukrainian SSR (in Russian), Ser.~A,
        {\bf 11}, 74 (1981).

\bibitem{KSS-85}
Yu.~Kagan, B.V.~Svistunov, and G.V.~Shlyapnikov,
% ``Effect of Bose condensation on inelastic processes in gases,''
JETP Lett.\ {\bf 42}, 209 (1985).

\bibitem{Burt97}
E.A.~Burt, R.W.~Ghrist, C.J.~Myatt, M.J.~Holland,
                E.A.~Cornell, and C.E.~Wieman,
% ``Coherence, Correlations, and Collisions: What One Learns
%         about Bose-Einstein Condensates from Their Decay,''
Phys.\ Rev.\ Lett.\ {\bf 79}, 337 (1997).

\bibitem{Petrov-octs}
D.~Petrov,
% ``Three-boson problem near a narrow Feshbach resonance'',
talk at the Workshop on Strongly Interacting Quantum Gases,
Ohio State University, April 2005.
%( http://octs.osu.edu/images/Gases/Talks/Petrov.pdf ).

\bibitem{MOG05}
J.H.~Macek, S.~Ovchinnikov, and G.~Gasaneo,
%``Exact solution for boson-diboson elastic scattering at zero energy 
%	in the shape independent model,''
Phys.\ Rev.\ A {\bf 72}, 032709 (2005).

\bibitem{NM-99}
E.~Nielsen and J.H.~Macek,
% ``Low-energy Recombination of Identical Bosons by Three-Body Collisions,''
Phys.\ Rev.\ Lett.\ {\bf 83}, 1566 (1999).

\bibitem{EGB-99}
B.D.~Esry, C.H.~Greene, and J.P.~Burke,
% ``Recombination of Three Atoms in the Ultracold Limit,''
Phys.\ Rev.\ Lett.\ {\bf 83}, 1751 (1999).

\bibitem{BBH-00}
P.F.~Bedaque, E.~Braaten, and H.-W.~Hammer,
% ``Three-body recombination in Bose gases with large scattering length,''
Phys.\ Rev.\ Lett.\ {\bf 85}, 908 (2000).
%        [arXiv:cond-mat/0002365].

\bibitem{BH02}
E.~Braaten and H.-W.~Hammer,
% ``Universality in the Three-Body Problem for $^4$He Atoms,''
Phys.\ Rev.\ A {\bf 67}, 042706 (2003).
%        [arXiv:cond-mat/0203421].

\bibitem{SEGB02}
H.~Suno, B.D.~Esry, C.H.~Greene, and J.P.~Burke, Jr.,
% ``Three-body recombination of cold helium atoms,''
Phys.\ Rev.\ A {\bf 65}, 042725 (2002).

\bibitem{Braaten:2003yc}
E.~Braaten and H.-W.~Hammer,
% ``Enhanced Dimer Relaxation in an Atomic and Molecular BEC,''
Phys.\ Rev.\ A {\bf 70}, 042706 (2004).
%[arXiv:cond-mat/0303249].

\bibitem{NSE00}
E.~Nielsen, H.~Suno, and B.D.~Esry,
% ``Efimov resonances in atom-diatom scattering,''
Phys.\ Rev.\ A {\bf 66}, 012705 (2002).

\bibitem{BH01}
E.~Braaten and H.-W.~Hammer,
% ``Three-body recombination into deep bound states
  %       in a Bose gas with large scattering length,''
Phys.\ Rev.\ Lett.\  {\bf 87}, 160407 (2001).
%        [arXiv:cond-mat/0103331].

\bibitem{NFJ98}
E.~Nielsen, D.V.~Fedorov, and A.S.~Jensen, 
% ``The structure of the atomic helium trimers: halos and Efimov states,''
J.\ Phys.\ B {\bf 31}, 4085 (1998).

\bibitem{RY00}
V.~Roudnev and S.~Yakovlev, 
% ``Investigation of $^4$He$_3$ trimer on the base of 
 %        Faddeev equations in configuration space,''
Chem.\ Phys.\ Lett.\ {\bf 328}, 97 (2000).

\bibitem{MSSK01}
A.K.~Motovilov, W.~Sandhas, S.A.~Sofianos, and E.A.~Kolganova, 
% ``Binding energies and scattering observables 
 %        in the $^4$He$_3$ atomic system,''
Eur.\ Phys.\ J.\ D {\bf 13}, 33 (2001).

\bibitem{LMKGG}
F.~Luo, G.C.~McBane, G.~Kim, C.F.~Giese, and W.R.~Gentry, 
% ``The weakest bond: Experimental observation of helium dimer,''
J.\ Chem.\ Phys.\ {\bf 98}, 3564 (1993).

\bibitem{LGG96}
F.~Luo, C.F.~Giese, and W.R.~Gentry, 
% ``Direct measurement of the size of the helium dimer,''
J.\ Chem.\ Phys.\ {\bf 104}, 1151 (1996).

\bibitem{STo94}
W.~Sch\"ollkopf and J.P.~Toennies, 
Science {\bf 266}, 1345 (1994).

\bibitem{BST02}
L.W.~Bruch, W.~Sch\"ollkopf and J.P.~Toennies, 
J.\ Chem.\ Phys.\ {\bf 117}, 1544 (2002).

\bibitem{Grisenti}
R.E.~Grisenti, W.~Sch\"ollkopf, J.P.~Toennies, G.C.~Hegerfeldt, 
        T.~K\"ohler, and M.~Stoll,
% ``Determining the Bond Length and Binding Energy of the Helium Dimer
%        by Diffraction from a Transmission Grating,''
Phys.\ Rev.\ Lett.\ {\bf 85}, 2284 (2000).

\bibitem{LDD77}
T.K.~Lim, Sister K.~Duffy, and W.C.~Damert, 
% ``Efimov State in the $^4$He Trimer,''
Phys.\ Rev.\ Lett.\ {\bf 38}, 341 (1977).

\bibitem{BHK99}
P.F.~Bedaque, H.-W.~Hammer, and U.~van~Kolck,
% ``Renormalization of the three-body system with short-range interactions,''
Phys.\ Rev.\ Lett.\ {\bf 82}, 463 (1999).
%        [arXiv:nucl-th/9809025].

\bibitem{BHK99b}
P.F.~Bedaque, H.-W.~Hammer, and U.~van~Kolck,
% ``The Three-Boson System with Short-Range Interactions,''
Nucl.\ Phys.\ A {\bf 646}, 444 (1999). 
%        [arXiv:nucl-th/9811046].

\bibitem{Esry96a}
B.D.~Esry, C.D.~Lin, and C.H.~Greene,
% ``Adiabatic hyperspherical study of the helium trimer,''
Phys.\ Rev.\ A {\bf 54}, 394 (1996).

\bibitem{AJM95}
R.A.~Aziz, A.R.~Janzen, and M.R.~Moldover,
% ``Ab Initio Calculations for Helium: 
%         A Standard for Transport Property Measurements,''
Phys.\ Rev.\ Lett.\ {\bf 74}, 1586 (1995).

\bibitem{Petrov04}
D.S.~Petrov,
%``Three-boson problem near a narrow Feshbach resonance,''
Phys.\ Rev.\ Lett.\ {\bf 93}, 143201 (2004).
%[arXiv:cond-mat/0404036].

\bibitem{Roberts00}
J.L.~Roberts, N.R.~Claussen, S.L.~Cornish, and C.E.~Wieman,
% "Magnetic Field Dependence of Ultracold Inelastic Collisions near a
% Feshbach Resonance,"
Phys.\ Rev.\ Lett.\ {\bf 85}, 728 (2000).

\bibitem{Marte02}
A.~Marte, T.~Volz, J.~Schuster, S.~D\"urr, G.~Rempe, E.G.M.~van Kempen,
and B.J.~Verhaar,
% "Feshbach Resonances in Rubidium 87: Precision Measurement and Analysis,"
Phys.\ Rev.\ Lett.\ {\bf 89}, 283202 (2002).

\bibitem{Smirne06}
G.~Smirne, R.M.~Godun, D.~Cassettari, V.~Boyer, C.J.~Foot, T.~Volz, 
N.~Syassen, S.~D\"urr, G.~Rempe, M.D.~Lee, K.~Goral, and T.~K\"ohler,
% "Collisional relaxation of Feshbach molecules and three-body
% recombination in 87Rb Bose-Einstein condensates"
arXiv:cond-mat/0604183.

\bibitem{Weber03}
T.~Weber, J.~Herbig, M.~Mark, H.-C.~N\"agerl, and R.~Grimm,
% ``Three-Body Recombination at Large Scattering Lengths 
%         in an Ultracold Atomic Gas,''
Phys.\ Rev.\ Lett.\ {\bf 91}, 123201 (2003).

%Feshbach Resonances in Cs_2 -- Cs_2 scattering:
\bibitem{Chin05}
C.~Chin, T.~Kr\"amer, M.~Mark, J.~Herbig, P.~Waldenburger, H.-C.~N\"agerl, 
and R.~Grimm,
% "Observation of Feshbach-Like Resonances in Collisions between Ultracold
% Molecules,"
Phys.\ Rev.\ Lett.\ {\bf 94}, 123201 (2005).

\bibitem{FB18Santos}
R.~Grimm, talk at 
18th International Conference on Few-Body Problems in Physics, 
Santos, Brazil, August 2006.

\bibitem{BH06}
E.~Braaten and H.-W.~Hammer, 
% ``Resonant Dimer Relaxation in Cold Atoms with a Large Scattering Length,''
arXiv:cond-mat/0610116.

\bibitem{AN72}
R.D.~Amado and J.V.~Noble,
% ``Efimov's Effect: A New Pathology of Three-Particle Systems. II,''
Phys.\ Rev.\ D {\bf 5}, 1992 (1972).

\bibitem{Efimov72}
V.~Efimov,
% ``Level Spectrum of Three Resonantly Interacting Particles,''
Sov.\ Phys.\ JETP Lett.\ {\bf 16}, 34 (1972)
[ZhETF Pis.\ Red.\ {\bf 16}, 50 (1972)].

\bibitem{Efimov73}
V.~Efimov,
% ``Energy Levels of Three Resonantly-interacting Particles,''
Nucl.\ Phys.\ A {\bf 210}, 157 (1973).

\bibitem{BE75}
A.~Bulgac and V.~Efimov,
% ``Spin dependence of the level spectrum of three resonantly 
%        interacting particles,''
Sov.\ J.\ Nucl.\ Phys.\ {\bf 22}, 153 (1976)
        [Yad.\ Fiz.\ {\bf 22}, 296 (1975)].

\bibitem{OS79} 
Yu.N.~Ovchinnikov and I.M.~Sigal,  
% ``Number of Bound States of Three-Body Systems and Efimov's Effect,'' 
Ann.\ Phys.\ {\bf 123}, 274 (1979).

\bibitem{FRS79} 
A.C.~Fonseca, E.F.~Redish, and P.E.~Shanley,  
% ``Efimov Effect in an Analytically Solvable Model,'' 
Nucl.\ Phys.\ A {\bf 320}, 273 (1979).

%%%%%%%%%%%%%%%%%%%%%%%%%%%%%%%%%%%%%%%%%%%%%%%%%%
%    Effective-range corrections
%%%%%%%%%%%%%%%%%%%%%%%%%%%%%%%%%%%%%%%%%%%%%%%%%%

%\bibitem{XB}
%{\bf B.  Effective-range corrections}

\bibitem {Efimov93}
V.~Efimov, 
%``Effective interaction of three resonantly interacting particles
%        and the force range,''
Phys.\ Rev.\ C {\bf 47}, 1876 (1993).

\bibitem {Efimov91}
V.~Efimov, 
%``Force-range correction in the three-body problem: Application to 
%three-nucleon systems''
Phys.\ Rev.\ C {\bf 44}, 2303 (1991).

\bibitem{EfT85}
V.~Efimov and E.G.~Tkachenko,
%``Explanation of the Phillips-Line in the Three-Nucleon Problem,''
Phys.\ Lett.\ {\bf 157B}, 108 (1985).

\bibitem{HM-01b}
H.-W.~Hammer and T.~Mehen,
%``Range Corrections to Doublet S-Wave Neutron-Deuteron Scattering,''
Phys.\ Lett.\ B {\bf 516}, 353 (2001).
%        [arXiv:nucl-th/0105072].

\bibitem{Afnan:2003bs}
I.R.~Afnan and D.R.~Phillips,
%``The three-body problem with short-range forces: renormalized equations 
%        and regulator-independent results,''
 Phys.\ Rev.\ C {\bf 69}, 034010 (2004).
%        [arXiv:nucl-th/0312021].

\bibitem{BGHR03}
P.F.~Bedaque, H.W.~Grie{\ss}hammer, H.-W.~Hammer, and G.~Rupak,
%``Low Energy Expansion in the Three Body System to All Orders
%        and the Triton Channel,''
Nucl.\ Phys.\  A {\bf 714}, 589 (2003).
%        [arXiv:nucl-th/0207034].

\bibitem{Griesshammer:2005ga}
H.W.~Grie{\ss}hammer,
%``Naive Dimensional Analysis for Three-Body Forces Without Pions,''
Nucl.\ Phys.\ A {\bf 760}, 110 (2005).
%[arXiv:nucl-th/0502039].

\bibitem{Birse:2005pm}
M.C.~Birse,
%``Renormalisation-group analysis of repulsive three-body systems,''
J.\ Phys.\ A {\bf 39}, L49 (2006).
%[arXiv:nucl-th/0509031].

\bibitem{Platter:2006ev}
L.~Platter and D.R.~Phillips,
%``The Three-Boson System at Next-To-Next-To-Leading Order,''
Few-Body Syst.~(2006) [arXiv:cond-mat/0604255].

%\cite{Platter:2006ad}
\bibitem{Platter:2006ad}
  L.~Platter,
  %``The Three-Nucleon System at Next-To-Next-To-Leading Order,''
  Phys.\ Rev.\ C {\bf 74}, 037001 (2006).
%  [arXiv:nucl-th/0606006].
  %%CITATION = NUCL-TH 0606006;%%

%%%%%%%%%%%%%%%%%%%%%%%%%%%%%%%%%%%%%%%%%%%%%%%%%%
%    Scattering models
%%%%%%%%%%%%%%%%%%%%%%%%%%%%%%%%%%%%%%%%%%%%%%%%%%

%\bibitem{XD}
%{\bf D. Scattering models}

\bibitem{KMCWH02}
S.J.J.M.F.~Kokkelmans, J.N.~Milstein, M.L.~Chiofalo, R.~Walser, 
        and M.J.~Holland,
%``Resonance Superfluidity: Renormalization of Resonance Scattering Theory,''
Phys.\ Rev.\ A {\bf 65}, 053617 (2002).
%        [arXiv:cond-mat/0112283].


%%%%%%%%%%%%%%%%%%%%%%%%%%%%%%%%%%%%%%%%%%%%%%%%%%
%    The N-body problem for N > 4
%%%%%%%%%%%%%%%%%%%%%%%%%%%%%%%%%%%%%%%%%%%%%%%%%%

%\bibitem{XA}
%{\bf A. The $N$-body problem for $N \geq 4$}

\bibitem{Yakubovsky:1966ue}
O.A.~Yakubovsky,
%``On the Integral Equations in the Theory of N Particle Scattering,''
Sov.\ J.\ Nucl.\ Phys.\  {\bf 5}, 937 (1967)
        [Yad.\ Fiz.\  {\bf 5}, 1312 (1967)].

\bibitem{GrSa67}
P.~Grassberger and W.~Sandhas,
%``Systematical Treatment of the Non-Relativistic N-Particle Problem,''
Nucl.\ Phys.\ B {\bf 2}, 181 (1967).

\bibitem{AFG95}
S.K.~Adhikari, T.~Frederico, and I.D.~Goldman,
%``Perturbative Renormalization in Quantum Few-Body Problems,''
Phys.\ Rev.\ Lett.\ {\bf 74}, 487 (1995).

\bibitem{Lepage-pc}
G.P.~Lepage, unpublished.

\bibitem{Platter:2004qn}
L.~Platter, H.-W.~Hammer, and U.-G.~Mei{\ss}ner,
%``The Four-Boson System with Short-Range Interactions,''
Phys.\ Rev.\ A {\bf 70}, 052101 (2004).
%[arXiv:cond-mat/0404313].

\bibitem{BlGr00}
D.~Blume and C.H.~Greene,
%``Monte Carlo Hyperspherical Description of Helium Cluster Excited States,''
J.\ Chem.\ Phys.\ {\bf 112}, 8053 (2000).

\bibitem{Yamashita06}
M.T.~Yamashita, L.~Tomio, A.~Delfino, and T.~Frederico,
%``The four-boson scale near a Feshbach resonance,''
Europhys.\ Lett.\ {\bf 75}, 555 (2006).
%[arXiv:cond-mat/0602549].

\bibitem{dinc06}
J.P.~D'Incao and B.D.~Esry,
%''Enhancing the observability of the Efimov effect in ultracold atomic gas mixtures,''
Phys.\ Rev.\ A {\bf 73}, 030703(R) (2006).

\end{thebibliography}
\end{document}